\documentclass[preprint,12pt]{elsarticle}
\usepackage{lipsum}
\makeatletter
\def\ps@pprintTitle{%
	\let\@oddhead\@empty
	\let\@evenhead\@empty
	\def\@oddfoot{}%
	\let\@evenfoot\@oddfoot}
\makeatother

\usepackage{amsmath} 
\usepackage{amssymb}
\usepackage{booktabs}
\usepackage{color}

\begin{document}

\begin{frontmatter}

\title{Suppression of free convection effects for spherical 1~kg mass prototype}

\author[Label1]{Sebastian Sachs}
\author[Label2]{Thomas Fr\"ohlich}
\author[Label1]{J\"org Schumacher}

\address[Label1]{Institute of Thermodynamics and Fluid Mechanics, Technische Universit\"at Ilmenau, D-98684 Ilmenau, Germany}
\address[Label2]{Institute of Process Measurement and Sensor Technology, Technische Universit\"at Ilmenau, D-98684 Ilmenau, Germany}

\begin{abstract}
We investigate the free convection processes in the vicinity of a spherical 1 kg mass standard by two- and three-dimensional direct numerical simulations using a spectral element method. Our focus is on the determination and suppression of updraft forces in a high-precision mass comparator which are caused by temperature differences between mass standard and its environment in the millikelvin range -- a source of systematic uncertainties in the high-precison mass determination. A two-dimensional model is presented first, which obtains a good agreement with previous laboratory measurements for the smaller temperature differences up to 15 mK. The influence of different boundary conditions and side lengths of the square domain is discussed for the mass standard positioned in the center of the chamber. The complexity is increased subsequently in configurations with additional built-ins for counter heating in form of planar plates or hemispherical shells above the mass standard. The latter ones lead to a full compensation of the updraft force. Three-dimensional simulations in a closed cubic chamber confirm the two-dimensional findings and additionally reveal complex secondary flow pattern in the vicinity of the mass standard. The reduction of the heat transfer due to the built-ins is also demonstrated by a comparison of the Nusselt numbers as a function of the Rayleigh number in the chosen parameter range. Our simulations suggest that such additional constructive measures can enhance the precision of the mass determination by suppression of free convection and related systematic uncertainties.
\end{abstract}


\end{frontmatter}


\section{Introduction} \label{Intro}
\begin{figure}[h]
	\centering
	\includegraphics[scale=0.46]{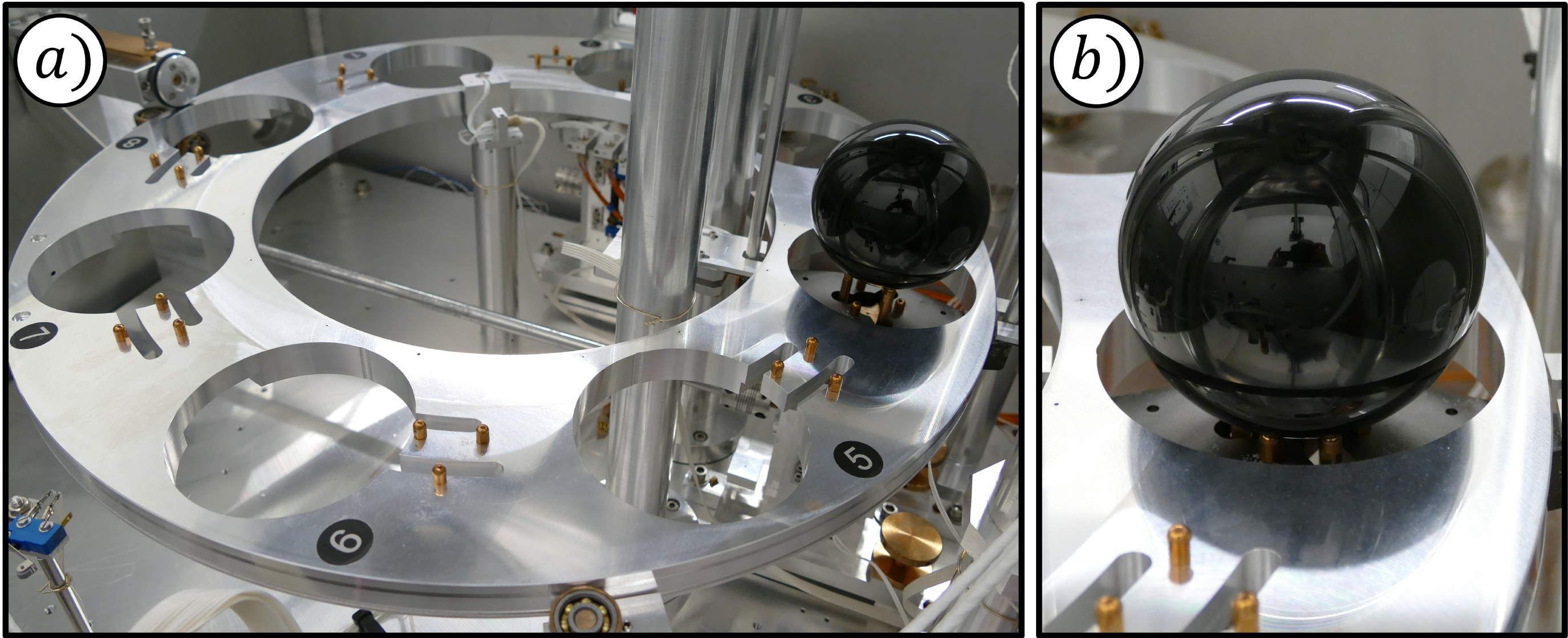}
	\caption{Measurement chamber of the mass comparator CLL1007 with a spherical mass standard and additional components (a). Close-up of the mass standard in weighing position (b).}
	\label{img:Massekomparator}
\end{figure}

Heat transfer and thermally induced convection processes can lead to systematic measurement uncertainties $\Delta m$ which is particularly relevant in high-precision weighing. Their systematic study requires investigations of the free convection dynamics in complex geometries with different boundary conditions. Such processes have been analyzed, both experimentally and numerically, in various studies. We mention here the first investigations by Gläser et al.~\cite{glaesser1990,glaesser1993,glaesser1999} and subsequently by Mana et al.~\cite{Mana2002} which evaluated the apparent mass differences in deviating measurement setups for mass artifacts with different weights and shapes. The influence of significantly lower temperature differences on measurements of spherical and cylindrical 1 kg mass standards in a high-precision mass comparator as shown in Fig. \ref{img:Massekomparator} was later studied by Schreiber et al. \cite{Schreiber2015}. Under realistic conditions with observed temperature differences of the order of a few millikelvin, the mass measurement result was affected in the microgram range. These investigations provide the motivation for our present numerical investigations. 

In this work, we want to consider a 1 kg mass prototype whose temperature is marginally higher than that of the surrounding air. Due to buoyancy forces in the immediate vicinity of the mass standard, a laminar boundary layer flow is formed which causes friction and pressure forces on the mass surface. The resulting updraft force $F=g\Delta m$ creates an apparent mass difference $\Delta m$ when comparing with a reference mass standard without a temperature difference, i.e., $\Delta T=0$. Our current work differs from and extends the study by Schreiber et al. \cite{Schreiber2015} in several ways. (1) Here, we use a highly accurate spectral element direct numerical simulation method to investigate the natural (or free) convection and the resulting forces. (2) The mass standard in real settings is typically rather a silicon sphere than a cylindrical prototype. (3) Even though the Rayleigh numbers of the free convection process remain small, the present simulation method allows us to resolve the fields and their derivatives in the thin boundary layers correctly in order to determine the updraft forces. (4) The parallelized numerical scheme allows for efficient parametric search as we will discuss in detail. (5) Furthermore, we will implement a targeted flow suppression through additional built-ins above the spherical mass standard for counter heating. The aim is to reduce the apparent mass difference and even demonstrate a complete compensation of the updraft forces acting on the measurement object by geometrical variations of these additional built-ins. We will use the following convention in this work: $\Delta m>0$ corresponds with an updraft force which implies that the mass standard is effectively lighter; for $\Delta m<0$ one obtains a downdraft force with a heavier sphere. 

Several other numerical studies on free convection in closed cavities of different shapes focussed on the specific structure of the triggered flows and the resulting magnitude of the heat transfer.  In this context, the works of Yoon et al. \cite{yoon2010}, Lee et al. \cite{Lee2019} and Chen \cite{Chen2010} should be emphasized. Our focus is here on the resulting integral forces. By means of the Gauss-Lobatto-Legendre theorem, we can determine such forces on the surface of the sphere as accurately as the polynomial order of the expansion functions is set on each spectral element and in each space dimension. In order to reduce the number of influence parameters, the flow problem is first considered in a two-dimensional (2d) model with less computational effort. Finally, we extend the study to the full three-dimensional (3d) state. Our study should be considered as a first step to systematically reveal possible reasons for the mass uncertainties which arise in a complex geometry of a mass comparator and which can be suppressed by taking additional constructive measures as the built-ins which will be described in detail further below.

The outline of our manuscript is as follows. In section \ref{simulation} we present the Boussinesq equations as the mathematical model for the calculation of free convection and discuss the numerical method as well as the mesh generation in the specific geometries. The effect of different boundary conditions,  domain sizes and built-ins on the apparent mass difference will be analyzed with two-dimensional calculations in section \ref{results}. The essential results on the basis of 3d simulations will be discussed in section \ref{3Dresults}. This offers the opportunity to critically examine the previous model assumptions and to illustrate three-dimensional phenomena, such as the formation of secondary flows which are absent in the 2d case. We conclude with a summary and an outlook in section~5.

\section{Numerical Simulation Model}
\label{simulation}
\subsection{Boussinesq equations of motion and dimensionless parameters}
\label{equations}
The calculation of the velocity $u_i$, the pressure deviation $p'$ from the hydrostatic equilibrium profile and the temperature difference $\theta=T-T_\infty$ to the surroundings is based on the numerical solution of the following dimensionless Boussinesq equations \cite{Schumacher2013},
\begin{eqnarray}
\frac{\partial \tilde{u}_{ i } }{\partial \tilde{x}_{i}  } &=& 0\,, \label{eqn:Bouss1}\\
\frac{\partial \tilde{u}_{i}}{\partial \tilde{t}} + \tilde{u}_{j}\frac{\partial \tilde{u}_{i}}{\partial \tilde{x}_{j}} &=& -\frac{\partial \tilde{p'}}{\partial \tilde{x}_{i}} + \sqrt{\frac{Pr}{Ra}} \frac{\partial^{2}\tilde{u}_{i}}{\partial \tilde{x}_{j}^{2}}+ \tilde{\theta} \delta_{iz} \,, \label{eqn:Bouss2}\\
\frac{\partial \tilde{\theta}}{\partial \tilde{t}} + \tilde{u}_{j}\frac{\partial\tilde{\theta}}{\partial \tilde{x}_{j}} &=& \frac{1}{\sqrt{Ra Pr}} \frac{\partial^{2} \tilde{\theta}}{\partial \tilde{x}_{j}^{2}} \,.\label{eqn:Bouss3}
\end{eqnarray}
The index notation $i,j=x,y,z$ is used in combination with the Einstein summation convention. Dimensionless quantities are expressed with a tilde. The diameter of the mass standard $d$, the free fall velocity $w_0=\sqrt{g\alpha\Delta T d}$ and the maximum temperature difference $\Delta T$ are used for normalization. Here, $g$ describes the acceleration due to gravity and $\alpha$ the coefficient of thermal expansion. By applying the Oberbeck-Boussinesq approximation buoyancy forces are involved in eq. \eqref{eqn:Bouss2} and thus require a coupled calculation of the temperature and velocity field. A detailed derivation is given for example in Chillà and Schumacher \cite{Chilla2012}. With the Kronecker symbol $\delta_{iz}$ the effect of the buoyancy forces is limited to the vertical direction. The remaining coefficients are determined by the dimensionless Rayleigh number~$Ra$ and Prandtl number~$Pr$. With the fluid parameters kinematic viscosity~$\nu$ and thermal diffusivity~$\kappa$ they can be expressed as follows.
\begin{align}
Ra = \frac{g\alpha \Delta T d^{3}}{\nu \kappa}, \qquad Pr = \frac{\nu}{\kappa} \label{eqn:Rayleigh}
\end{align}
Since the fluid properties do not change within the calculations, the Rayleigh number serves as a measure of the temperature difference and will be varied in the range $120\leq Ra \leq 4590$. The Prandtl number is set to the constant value $Pr=0.71$ for convection in air.

\subsection{Numerical method}
The coupled system of partial differential equations is numerically solved with the spectral element package Nek5000 \cite{Nek5000} with full time resolution to also study transients. The code uses the spectral element method, that combines the advantages of the finite element method for discretizing complex geometries with high spectral accuracy. Detailed information on the application of the method is given for example in Deville et al. \cite{Deville2002} and Scheel et al. \cite{Schumacher2013}. 

The computational domain is first divided into several elements that represent the specific geometry. Within each of these spectral elements, the solution is calculated using basis functions $\pi_i$ of order~$N$ at $(N+1)$ Gauß-Lobatto-Legendre (GLL) quadrature nodes $\xi_i$. On each element the three-dimensional quantities are expressed by the following extension in form of a tensor product formulation on the reference element interval $\Lambda=[-1,1]^3$ \cite{Schumacher2013}.
\begin{align}
\boldsymbol{u}^e_N(x,y,z)=\sum_{i=0}^{N}\sum_{j=0}^{N}\sum_{k=0}^{N}\boldsymbol{u}(\xi_i,\xi_j,\xi_k)\pi_i(x)\otimes\pi_j(y)\otimes\pi_k(z)
\end{align}
Since the basis functions are known by Lagrange interpolation polynomials at the beginning of the calculation, only the basis coefficients $\boldsymbol{u}_{i,j,k}=\boldsymbol{u}(\xi_i,\xi_j,\xi_k)$ have to be evaluated at the corresponding GLL nodes. In order to calculate the apparent mass difference, both the pressure deviation $p'$ and the shear stress $\tau_w$ distribution on the surface $A_S$ of the mass standard must be integrated. These integrations translate into a finite weighted sum over all GLL nodes on the surface by taking the GLL weights into account. An exemplary calculation of the vertical component of the friction force $F_{\tau,z}$ in the spherical coordinates ($r,\theta,\phi$) is given by
\begin{align}
F_{\tau,z}= \int_{A_S}\tau_{w,z}\bigg(r=\frac{d}{2},\theta,\phi\bigg)~dA\approx \sum_{e=0}^{E_S}\sum_{j=0}^{N}\sum_{k=0}^{N} M^e_{i=0,j,k}\tau^e_{w,z}(\xi_0,\xi_j,\xi_k)\,.  \label{eqn:Ftauz}
\end{align}
The index $i=0$ captures the faces of the elements that correspond to the surface of the sphere. Note also that the vertically acting component of the wall shear stress $\tau_{w,z}=-\tau_w\sin{\theta}$ is used. The summation is evaluated at the surface elements $E_S$ of the sphere with $r=d/2$ and thus for the inner GLL nodes in radial direction. The GLL weights are summarized together with determinants of the Jacobian, which occur due to the mapping of the curved elements, in the element mass matrix $M^e_{i,j,k}$ \cite{Schumacher2013}. As a result, the calculation of the systematic measurement uncertainty can be implemented with spectral accuracy here.

\subsection{Construction of the spectral element mesh}
\label{mesh}
\begin{figure}[h]
	\centering
	\includegraphics[scale=0.46]{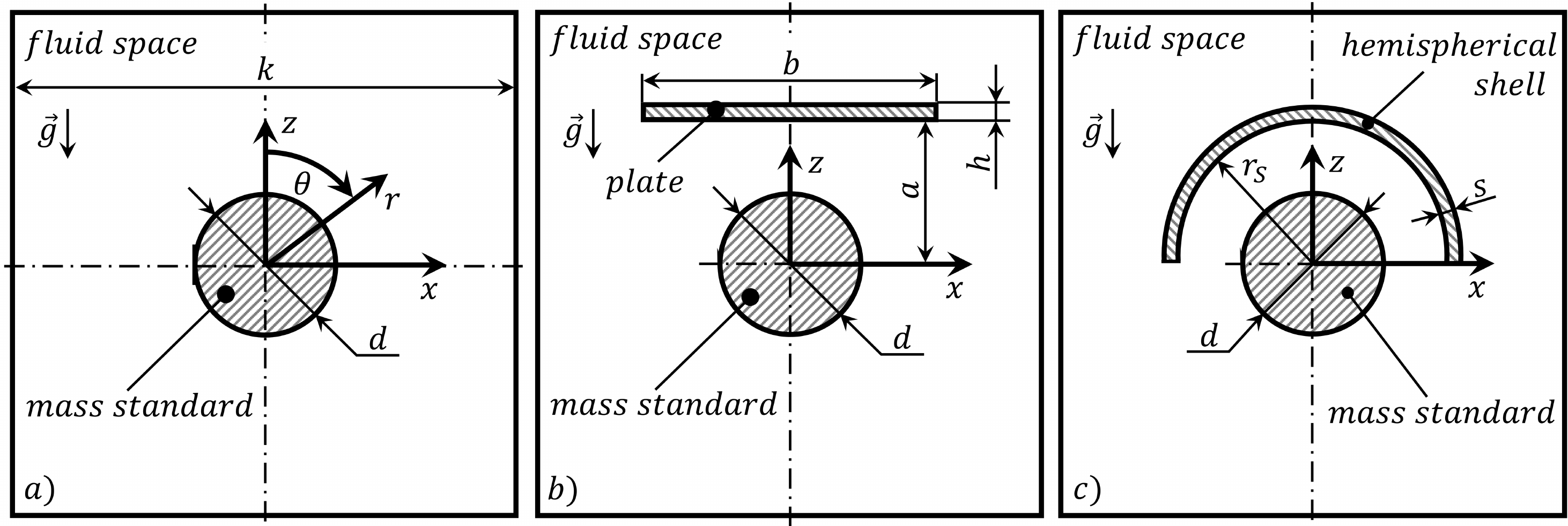}
	\caption{Sketch of the configuration with the mass standard in a cube-shaped domain (a). The flow suppression is performed either by an additional planar baffle plate (b) or a hemispherical shell (c).}
	\label{img:Kugel}
\end{figure}

In a simplified setup, the spherical mass standard is freely positioned in the center of a cube-shaped computational domain to calculate the unimpeded convection as shown in Fig. \ref{img:Kugel}a. The influences of the additional components such as the lifting device, an annular platform and receiving pins are neglected. While the diameter $d$ of the measuring object is constant, the edge length of the chamber $k$ is varied in section \ref{SIZE}. Subsequent calculations consider the targeted flow suppression by a straight baffle plate (see Fig. \ref{img:Kugel}b) and a hemispherical shell (see Fig. \ref{img:Kugel}c) above the mass prototype. The width $b=4d$ and height $h=0.1d$ of the baffle plate as well as the wall thickness $s=0.1d$ of the hemispherical shell remain unchanged in all simulations. The geometrical variation parameters, however, include the plate spacing $a$ and the radius $r_{S}$.

The spectral element mesh which covers the computational domain for the two-dimensional case is shown in Fig. \ref{img:Mesh}a. On the basis of a sensitivity analysis (see appendix A) of the grid we set the number of elements to 3200 with a polynomial order of the basis functions of $N = 7$. The mass prototype is modeled as a free space with no-slip boundary conditions and a constant temperature of $\tilde{\theta}(r=d/2) = 1.0$.  Boundary conditions on the outside of the flow domain will be varied in section \ref{BC}. The planar convection case assumes homogeneity in y-direction and can only approximate a full three-dimensional flow around the sphere. An axisymmetric convection flow setup would be a natural alternative that leaves however a thin free wedge in the mesh below the sphere to match spectral elements with different orientation. The axisymmetric case will thus not be considered here. Despite the expected deviations, the two-dimensional calculations enable a basic characterization of the occurring effects in combination with a systematic variation of several parameters.

A similar distribution of elements is achieved in the coordinate planes for the three-dimensional grid as shown in Fig. \ref{img:Mesh}b. With a maximum relative deviation of $0.02~\%$ in the updraft force in comparison to finer meshes, 15360 elements with a polynomial order of $N=7$ are used in the spatial grid. Further details are given in appendix B.
\begin{figure}[h]
	\centering
	\includegraphics[scale=0.42]{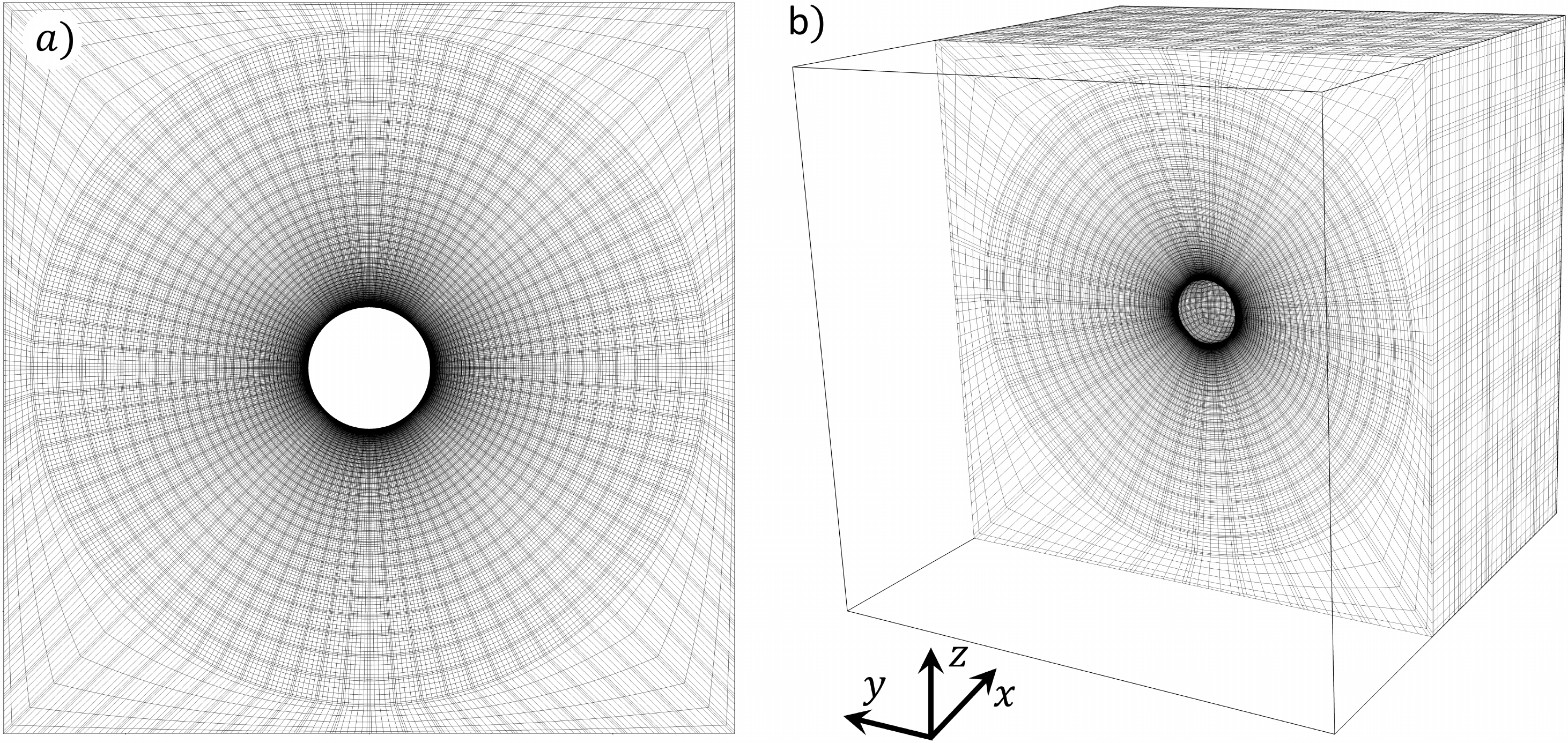}
	\caption{Planar (a) and spatial (b) spectral element grid. The three-dimensional grid is only visualized for the positive $x$-range for a better visibility.}
	\label{img:Mesh}
\end{figure}

\section{Two-dimensional analysis}
\label{results}
\subsection{Role of outer velocity boundary conditions on boundary layers}
\label{BC}
We start with a discussion of the flow behaviour for different boundary conditions at the four outer edges of the two-dimensional computational domain. Beside a free in- and outflow condition for all velocity components, $\partial u_i/\partial n=0$ with the normal direction $n$, the no-slip boundary conditions, $u_i=0$, are taken. The temperature specification at the outer walls accords to Dirichlet conditions, $\tilde{\theta}=0$. The edge length amounts to $\tilde{k}=12$ and the Rayleigh number is changed in the range $120\leq Ra\leq4590$, which equals a temperature difference of 1.45 mK $\leq \Delta T \leq$ 55.27 mK. This range of temperature differences agrees with the typical values in the high-precision mass comparator \cite{Schreiber2015}. After each analysis of the flow structures and boundary layers, the pressure and wall shear stress distribution is evaluated to calculate the friction and pressure component of the resulting updraft force.

\begin{figure}[h]
	\centering
	\includegraphics[scale=0.55]{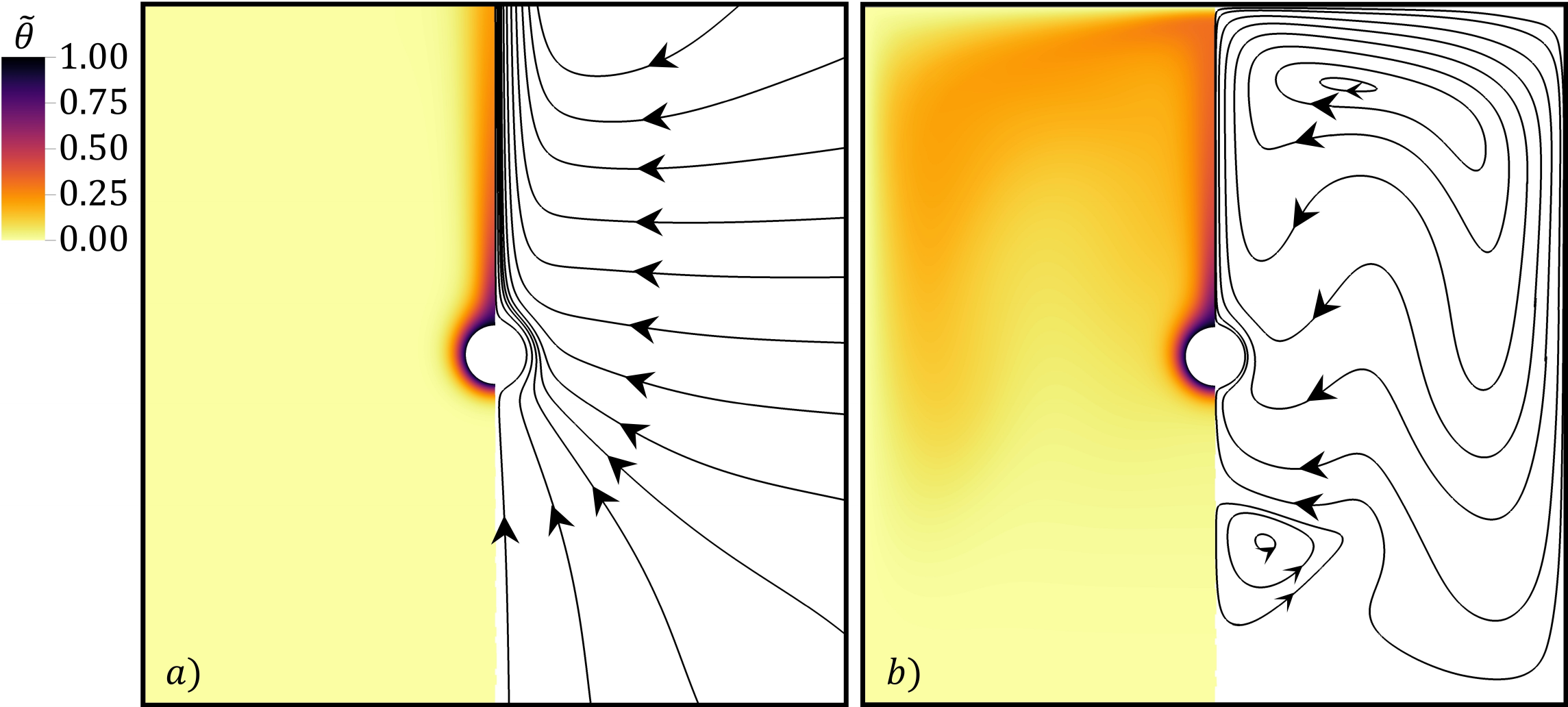}
	\caption{Isocontours (left halves) and streamlines (right halves) at a Rayleigh number $Ra=4590$. Free outflow (a) and no-slip (b) boundary conditions are chosen for the velocity field. The temperature is kept at $\tilde{\theta}=0$ at all outer boundaries. The dimensionless domain size $\tilde{k}=12$ in units of the diameter of the spherical mass standard.}
	\label{img:unbeeinBC}
\end{figure}

The flow structures are illustrated on the basis of the isocontours and streamlines in Fig. \ref{img:unbeeinBC}. Typical steady-state flow conditions are shown, which arise symmetrically to the vertical axis after a free-fall time of $\tilde{t}=200$ for a Rayleigh number of $Ra=4590$. The temperature increases in the vicinity of the mass standard in both constellations. Due to the presence of the gravitational acceleration $g$, buoyancy forces act on the fluid particles with lower mass density and provide the driving for free convection in the boundary layer.

A central flow column is formed above the spherical mass prototype. The jet-shaped outflow accelerates further gas from the sides and increases the velocity in flow direction transiently. A steady flow configuration is established after a finite period. In contrast to the free outflow case in Fig.~\ref{img:unbeeinBC}a, deflections occur at the borders of the computational domain in combination with recirculations, see  Fig.~\ref{img:unbeeinBC}b. These can affect the inflow to the mass standard and thus also the formation of the boundary layers. In addition to the main recirculation vortex in the upper area, a secondary vortex with significantly lower fluid velocities is formed below the mass prototype. This occurs in combination with deformed streamlines in the surroundings. The isocontours are dominated by convection and thus further extended above the sphere. Due to recirculation, fluid is transported back to the mass standard at a higher temperature.
\begin{figure}[h]
	\centering
	\includegraphics[scale=0.46]{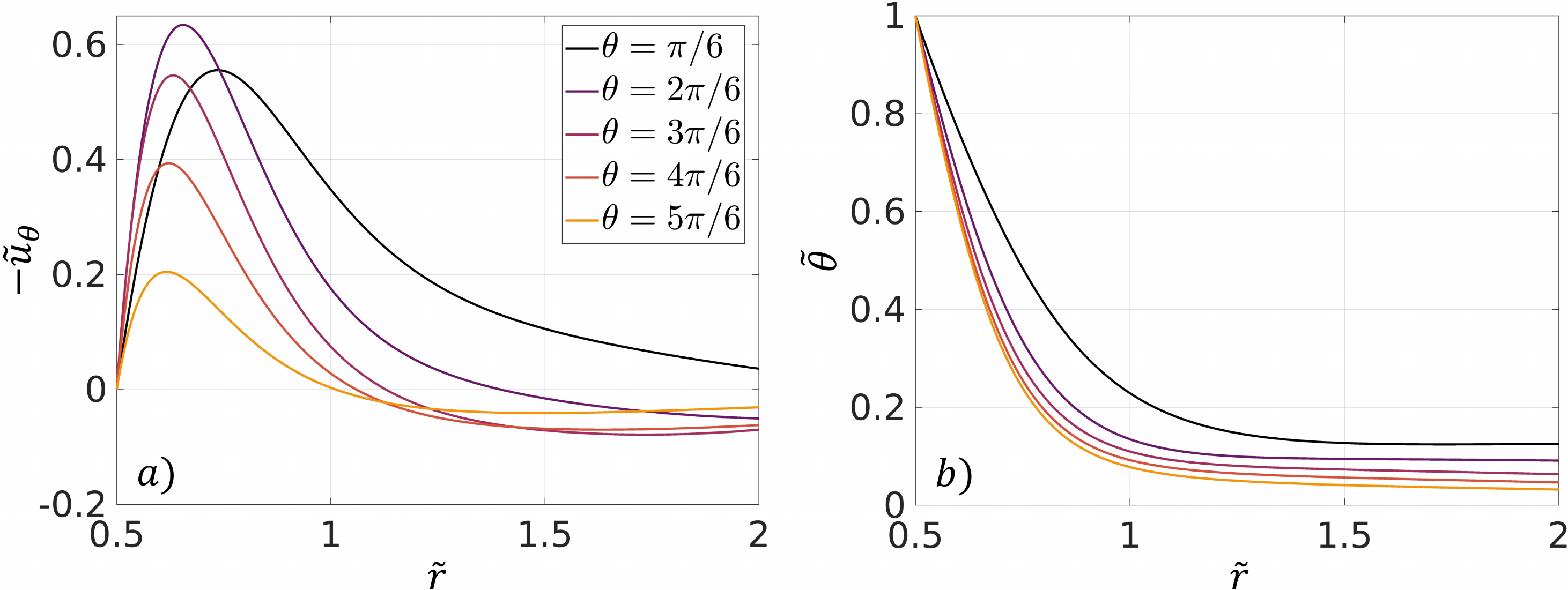}
	\caption{Boundary layer profiles of the polar velocity (the tangential velocity component) in panel (a) and the temperature in (b). All profiles are taken along axes  normal to the spherical surface. The polar angles $\theta$ are indicated in the legend which holds for both panels. No-slip conditions for the velocity field hold at the outer domain boundary.}
	\label{img:GSProfile}
\end{figure}
\begin{figure}[h]
	\centering
	\includegraphics[scale=0.46]{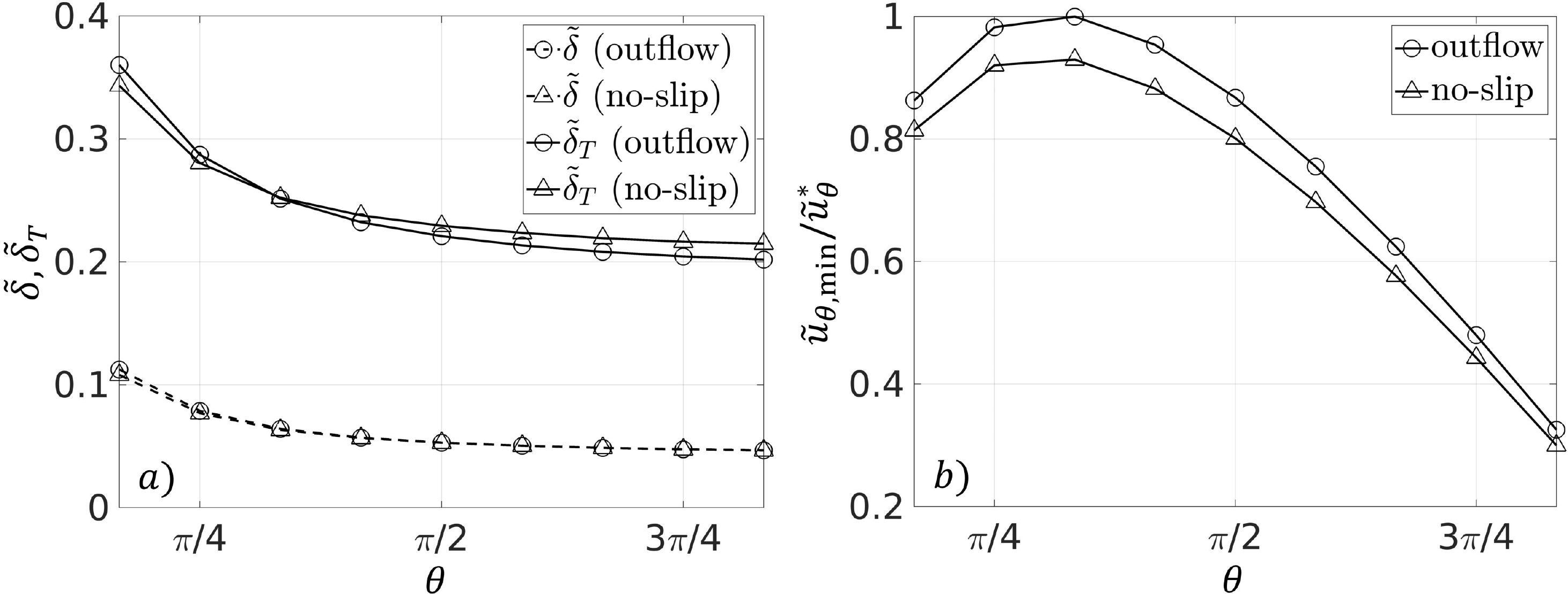}
	\caption{(a) Velocity and thermal boundary layer thickness for the free outflow and no-slip boundary conditions versus the polar angle $\theta$. (b) Minimum velocity $\tilde{u}_{\theta,{\rm min}}$ normalized with the global minimum value $\tilde{u}^{\ast}_{\theta}$ for the free outflow condition case as a function of the angle $\theta$. Data are for $Ra=4590$, at time $\tilde{t}=200$, and for a domain size of $\tilde{k}=12$.}
	\label{img:RBGS}
\end{figure}

The profiles of temperature and tangential velocity at different positions around the sphere are shown in Fig. \ref{img:GSProfile} for the highest Rayleigh number case of $Ra=4590$.  The temperature field forms a characteristic laminar boundary layer, the laminar velocity boundary layer passes through a minimum before increasing again close to zero. The maximum of $|u_{\theta}|$ increases steadily starting from the south pole into the northern hemisphere up to $\theta\le \pi/3$. The slope method, which is also applied in turbulent free convection \cite{Shishkina2009,Puits2010,Ovsyannikov2016,Ching2019}, quantifies the dimensionless temperature and velocity boundary layer thicknesses,  $\tilde{\delta}_T$ and $\tilde{\delta}$, at the spherical mass standard at different polar angles $\theta$. For this purpose, a tangent to the corresponding profile is taken at the point $\tilde{r}=1/2$. The intersection of this tangent with a horizontal tangent to the first (local) minimum of the corresponding profile quantifies the boundary layer thickness according to eq. \eqref{eqn:GSD2}. Regarding the velocity, the component $\tilde{u}_\theta(\tilde{r})$ in the direction of the polar angle $\theta$ is the tangential velocity. Note that we have $u_{\theta}<0$ as the polar angle starts from the north pole. The calculated boundary layer thicknesses are plotted in Fig. \ref{img:RBGS}a as a function of $\theta$. They are given by
\begin{equation}
\tilde{\delta} = \frac{\tilde{u}_{\theta,{\rm min}} }{\dfrac{\partial \tilde{u}_\theta}{\partial \tilde{r}}\Bigg|_{\tilde{r}=1/2}} \quad\mbox{and}\quad
\tilde{\delta}_T = \frac{1}{\dfrac{\partial \tilde{\theta}}{\partial \tilde{r}}\Bigg|_{\tilde{r}=1/2}}\bigg[\tilde{\theta}_{\rm min}-\tilde{\theta}\bigg(\tilde{r}=\frac{1}{2}\bigg)\bigg] \,.\label{eqn:GSD2}
\end{equation}

The profiles of both boundary layers in Fig. \ref{img:GSProfile} show that the evolving boundary layer close to the south pole in the southern hemisphere have the smallest thickness in this comparison (see the profiles at $\theta=5\pi/6$). The thickness of both laminar boundary layers grows steadily up to the detachment point  in the north pole region. It is also observable that the velocity boundary layer formation is insensitive to the conditions which are applied for the velocity field at the outer boundary of the computational domain. The thermal boundary layer is slightly affected in the southern hemisphere. In the no-slip case, the thickness is slightly larger which is in line with a somewhat smaller maximum velocity magnitude as shown in panel (b) of  Fig. \ref{img:RBGS}; it turns to smaller values at $\theta=\pi/3$.  The latter could be caused by the primary recirculation vortex above the mass standard.    

\begin{figure}[h]
	\centering
	\includegraphics[scale=0.46]{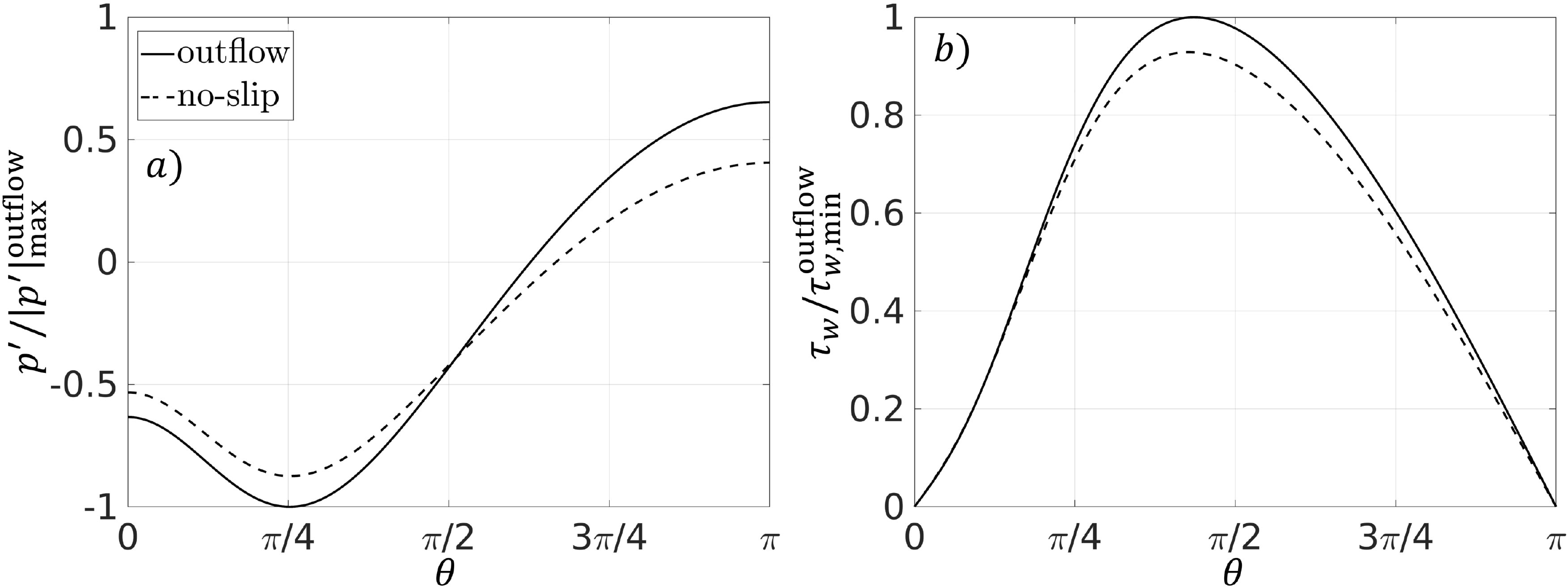}
	\caption{Distribution of pressure deviation (a) and wall shear stress ratios (b) at the mass standard surface as a function of the polar angle $\theta$. Pressure deviation data are normalized with the global maximum amplitude obtained with free outflow boundary conditions, wall shear stresses are normalized by the global minimum of the free outflow case.
This normalization is used for comparison of both runs.}
	\label{img:RBschub}
\end{figure}

The boundary layer thickness and fluid velocity have a strong impact on the shear stress $\tau_w$ and pressure $p'$ distribution at the spherical surface as shown in Fig.~\ref{img:RBschub}. The wall shear stress $\tau_w$ is given by
\begin{align}
\tau_w=\eta \frac{\partial u_\theta}{\partial r}\bigg|_{r=d/2} \,,
\label{eqn:newtonian}
\end{align}
with the dynamic viscosity $\eta=\nu\rho$ and the fluid density $\rho$. The corresponding wall shear stress ratio in panel (b) of Fig. \ref{img:RBschub} reaches its maximum close to the equatorial plane ($\theta=\pi/2$), while it converges to zero at the poles. The north pole is thus the detachment point with $\partial u_\theta/\partial r\approx 0$. Due to the steeper velocity gradient $\partial u_\theta/\partial r$ at the wall, the wall shear stress magnitude increases for outflow boundary conditions. The position of the maxima of the curves are found at nearly the same polar angle and are thus again independent of the outer velocity boundary conditions. 

There is a decrease of the pressure deviation starting from the south pole ($\theta=\pi$) where a stagnation point forms. This implies fluid is here accelerated up to the minimum at $\theta\approx \pi/4$. The pressure deviation profiles increase again for $\theta< \pi/4$ and reach a local maximum in the detachment region at the north pole. This increase is caused by the collision of the fluid that ascends around the sphere and is in line with an increasing velocity boundary thickness. Free outflow conditions lead again to slightly larger amplitudes and variations along the circumference as visible in panel (a) of Fig.~\ref{img:RBschub}.
\begin{figure}[h]
	\centering
	\includegraphics[scale=0.38]{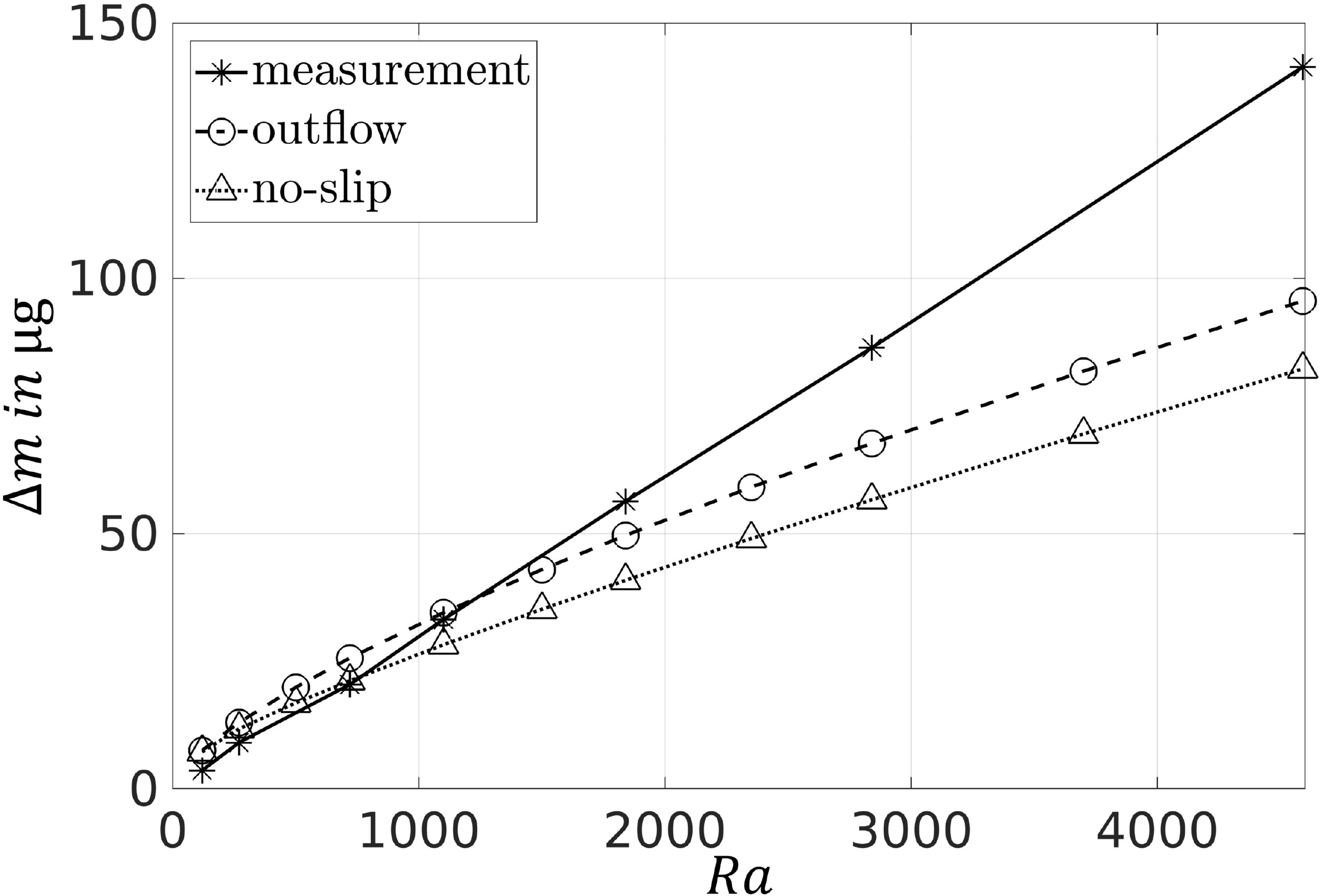}
	\caption{Course of the measured values \cite{Schreiber2015} and the numerical results for outflow and no-slip boundary conditions over the Rayleigh number with an edge length of $\tilde{k}=12$ and a free-fall time of $\tilde{t}=200$.}
	\label{img:RBAufwindRa}
\end{figure}

The effects on the apparent mass difference are discussed as a function of the Rayleigh number in Fig.~\ref{img:RBAufwindRa}. To calculate the friction and pressure forces, axisymmetry was assumed for the velocity and pressure deviation field. With an increasing Rayleigh number both boundary layers become thinner and the velocity magnitude increases. As a consequence, an increase of the resulting updraft forces becomes apparent. The positive values symbolize a mass reduction of the heated mass standard.

In relation to the measured values by Schreiber et al \cite{Schreiber2015}, a good agreement is achieved in the range of small temperature differences, especially when using the no-slip boundary condition. For larger Rayleigh numbers, however, the free outflow boundary condition is a slightly better approximation. The overall greater values in comparison to the no-slip case can be attributed to a higher pressure and frictional component of the updraft force. At the same time, the deviations of both simulation series results from the measurements in the balance increase with increasing Rayleigh number. Several possible reasons for these differences can be given. They comprise the three-dimensional flow character,  neglected components in comparison to the real mass comparator which will additionally affect the flow conditions, or existing measurement uncertainties (see also ref. \cite{Schreiber2015}).

As the numerical results demonstrate, the variation of the velocity conditions at the outer domain boundary do not affect the results qualitatively. Also the quantitative differences remain small with magnitudes $\Delta m \le 14\%$. Since the real measurements take place in a closed chamber made of aluminium and with finite dimensions, an interaction with the boundaries should not be excluded. As a consequence, the results with no-slip boundary conditions and a constant temperature specification of $\tilde{\theta}=0$ represent the best approximation and come closest to the real situation. These conditions remain fixed in the subsequent parameter studies which we will discuss in the following.

\subsection{Dependence on chamber size}
\label{SIZE}
To investigate the influence of the chamber size, the quadratic shape of the computational domain is retained, while the edge length is changed in the range $2\leq \tilde{k} \leq 12$. At all chamber walls, no-slip boundary conditions for the velocity field and constant temperature conditions, $\tilde{\theta} = 0$, are applied.
\begin{figure}[h]
	\centering
	\includegraphics[scale=0.34]{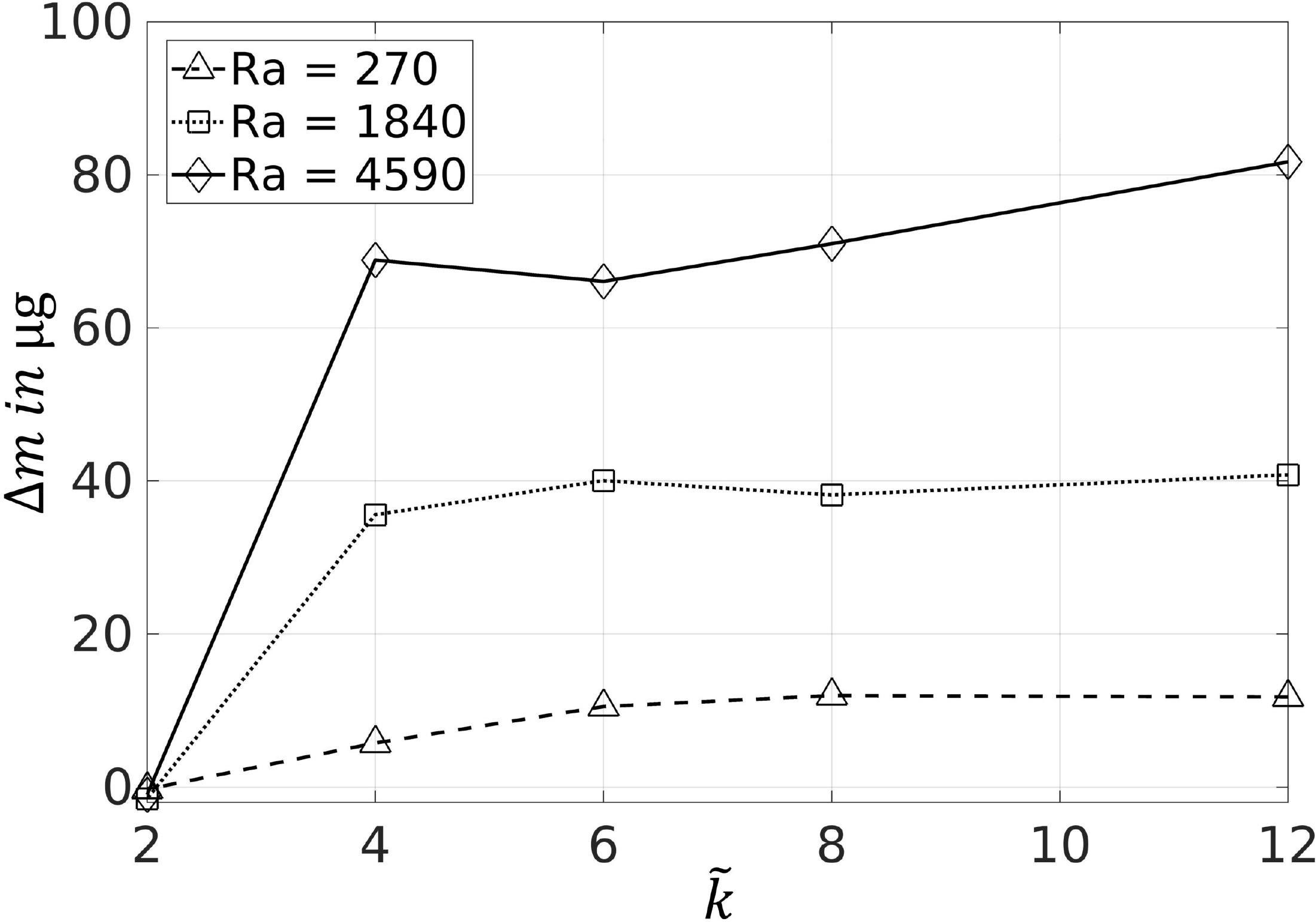}
	\caption{Change of the mass difference $\Delta m$ for three representative Rayleigh numbers $Ra$ (see the legend) plotted over the edge length $\tilde{k}=k/d$ which is given in units of the sphere diameter $d$.}
	\label{img:BoxAufwind}
\end{figure}

The curves of the resulting mass difference are plotted in Fig. \ref{img:BoxAufwind} for three different Rayleigh numbers versus the dimensionless edge length $\tilde{k}$. In the case of $\tilde{k} = 2$ the distance between the mass standard and the upper rigid boundary of the computational domain is so small that the pressure deviation distribution on the surface is dominated by the pressure increase due to the flow deflection point at the top boundary. As a result, a pressure force acts in the negative $z$-direction and compensates the frictional component of the resulting updraft force. The mass difference is thus practically zero. The figure also shows that the differences first grow for $\tilde{k}>2$ with chamber size for all three Rayleigh numbers.  In accordance with the results from section~\ref{BC}, larger systematic measurement uncertainties $\Delta m$ are found for greater temperature differences $\Delta T$ or higher Rayleigh numbers. For $\tilde{k} \gtrsim 4$ the susceptibility with respect to the domain size becomes much smaller. The deviations remain nearly constant or grow very weakly only for the largest Rayleigh number case.  The free convection dynamics around the sphere is now basically independent of domain size. 

Taking into account the real measuring arrangement in the mass comparator experiment, an edge length of $\tilde{k}=8$ is implemented in all further simulations. This specification offers a good approximation to the design of the measuring cell with moderate deviations of the updraft force to calculations with a larger fluid space. 

\subsection{Flow suppression by additional built-ins for counter heating}
\label{Plate}
In the following the targeted flow suppression by an additional baffle plate and a hemispherical shell is analyzed according to the description in section~\ref{mesh}. The aim of this modification is to reduce the systematic measurement uncertainty $\Delta m$ by considering two mechanisms. On the one hand, the flow deflection at the obstacles causes an increase in pressure above the mass standard, which influences the pressure distribution at the spherical surface. On the other hand, free natural convection is disturbed, which may reduce the frictional component of the updraft force. 

\begin{figure}[h]
	\centering
	\includegraphics[scale=0.47]{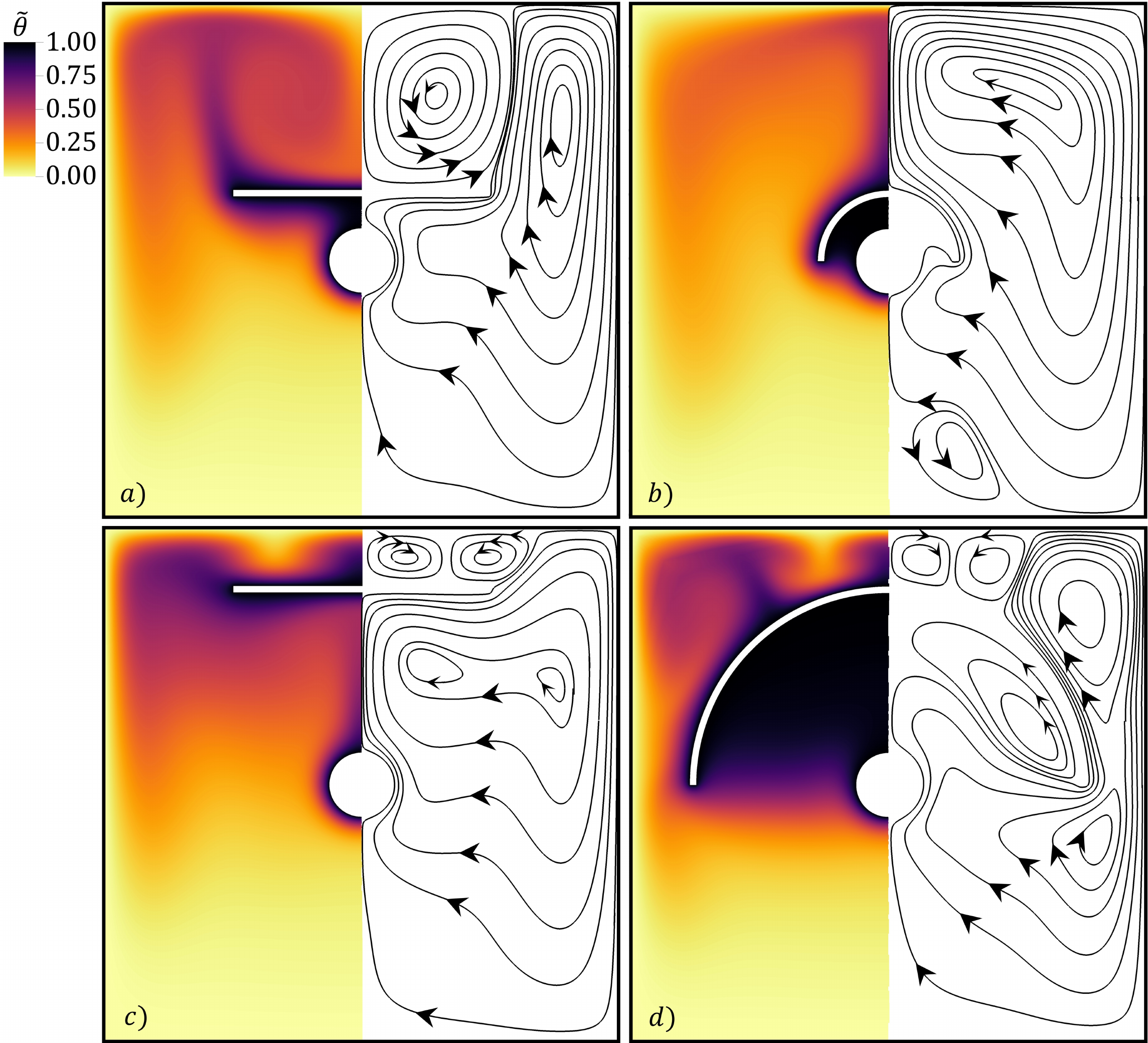}
	\caption{Isocontours (left halves) and streamlines (right halves) at a Rayleigh number $Ra=4590$. Straight baffle plate (a,c) and hemispherical shell (b,d) are chosen for the counter heating by built-ins. The temperature is kept at $\tilde{\theta}=0$ at all outer boundaries. The dimensionless domain size $\tilde{k}=8$ in units of the diameter of the spherical mass standard. In panels (a) and (c), we take $\tilde{a}=a/d=1$ and $\tilde{a}=3$, respectively. The temperature $\tilde{\theta}_P=1$. In panels (b) and (d), we take $\tilde{r}_s=r_s/d=1$ and $\tilde{r}_s=3$, respectively. The temperature is also $\tilde{\theta}_S=1$.}
	\label{img:PPStream}
\end{figure}

No-slip and constant temperature boundary conditions are implemented again at the obstacles for the velocity and temperature fields. Figure \ref{img:PPStream} illustrates the resulting flow behavior and temperature isocontours. The imposed temperature of the built-ins is here the same as the one on the surface of the mass standard. For example, clearly visible is the formation of convection rolls in the zone between the plate and the top wall in panel (c) of the figure. The visualizations demonstrate clearly that the structure of both fields depends sensitively on the geometry of these built-ins. 

\begin{figure}[h]
	\centering
	\includegraphics[scale=0.44]{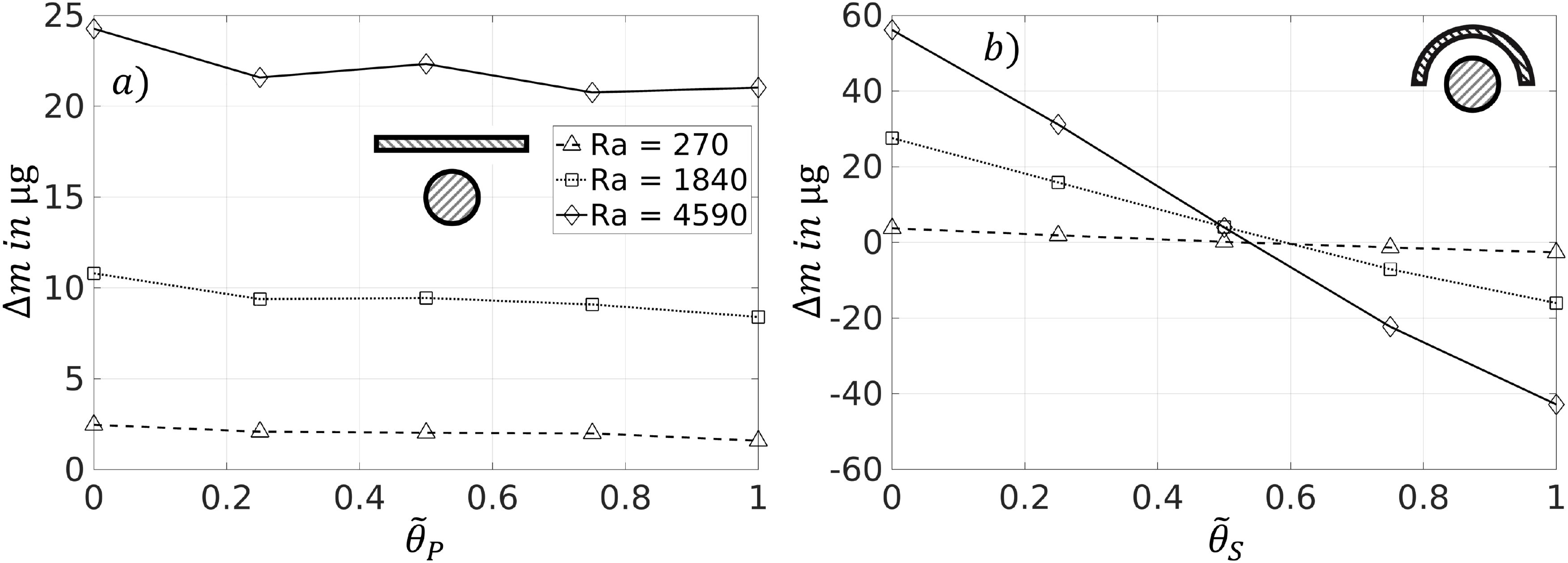}
	\caption{Development of the apparent mass difference as a function of the surface temperature of the straight baffle plate (a) and the hemispherical shell (b) for a plate distance $\tilde{a}=1$, a radius $\tilde{r}_s=2$ and various Rayleigh numbers.}
	\label{img:BeeinTemp}
\end{figure}

The effects of different surface temperatures at the baffle plate $\tilde{\theta}_P$ and the hemispherical shell $\tilde{\theta}_S$ on the apparent mass difference are summarized in Fig. \ref{img:BeeinTemp} for three different Rayleigh numbers. If the temperature (or in other words the counter heating) remains below the temperature of the mass prototype, a part of the supplied heat is transferred to the obstacles. This has consequences for the organization of the flow. While the fluid movement for low temperatures is mainly located below the obstacles, Rayleigh-Bénard convection rolls form in the upper region of the computational domain at higher obstacle temperatures $\tilde{\theta}_P$. The point at which Rayleigh-Bénard convection starts above the baffle plate is exactly determined by  
\begin{align}
Ra^{\ast}= \dfrac{g\alpha \theta_p \left[\frac{k}{2}-(a+h)\right]^3}{\nu\kappa} \gtrsim Ra_c = 1708 \,,
\label{eqn:newtonian}
\end{align}
when using the quantities in Fig. \ref{img:Kugel} and eq. \eqref{eqn:Rayleigh}. A similar (but not the same) estimate would follow for the case with a hemispherical obstacle. For temperatures $\tilde{\theta}_P>0$ or $\tilde{\theta}_S>0$ the obstacle thus represents an additional heating source.

The constrained fluid motion and the imposed pressure gradient manifest themselves in a reduction of the updraft force by at least $65~\%$ when using the straight baffle plate as displayed in panel (a) of Fig. \ref{img:BeeinTemp}. Also, $\Delta m$ is slightly decreasing for all three $Ra$ with increasing $\tilde{\theta}_P$.  In the case of the hemispherical shell, the system reacts much more sensitively to a change in temperature~$\tilde{\theta}_S$ as displayed in panel (b) of the same figure. This can be attributed to a homogenization of the fluid temperature and thus also fluid density below the spherical obstacle with increasing surface temperature. As a consequence, there is a stronger deceleration of the flow and an increase in pressure, which dominates the pressure distribution on the surface of the mass prototype. For increasing Rayleigh number, these effects manifest in steeper decaying curves.
\begin{figure}[h]
	\centering
	\includegraphics[scale=0.44]{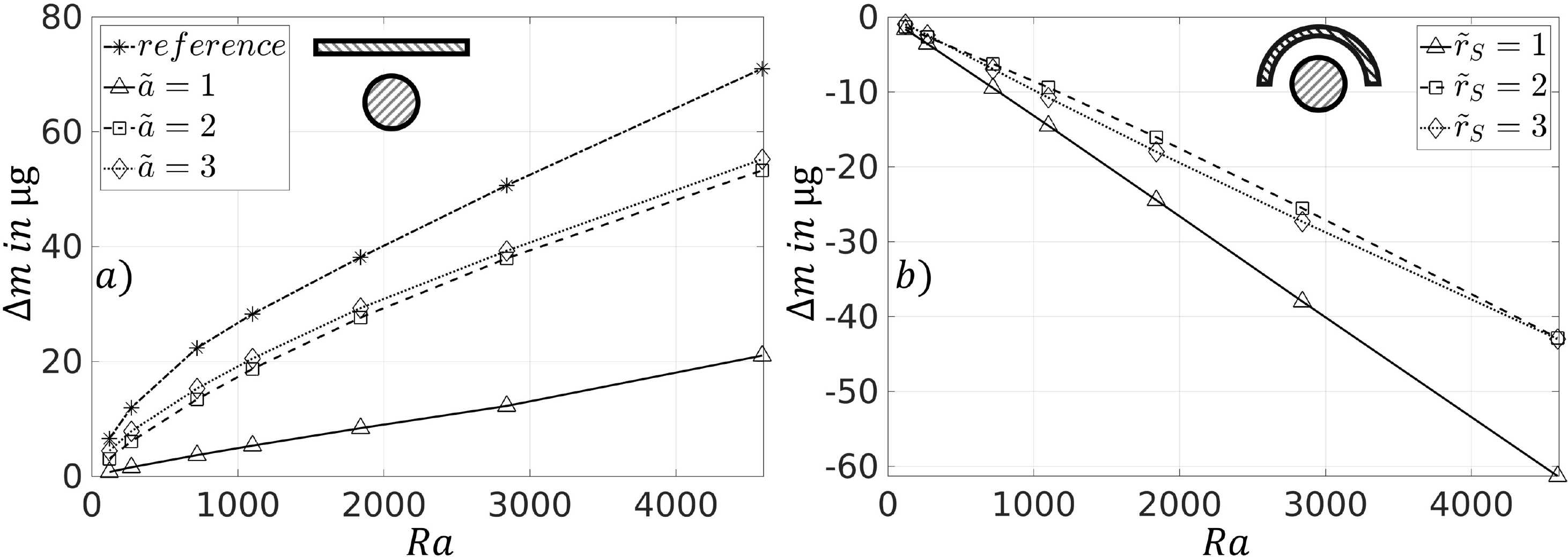}
	\caption{Numerical results on the influence of the plate distance $\tilde{a}$ (a) and the radius~$\tilde{r}_S$ of the hemispherical shell (b) with the surface temperatures $\tilde{\theta}_P=\tilde{\theta}_S=1$. The reference case in (a) stands for the case without an obstacle (see legend). }
	\label{img:BeeinAbstand}
\end{figure}

Furthermore, a sign change within the variation interval occurs when using the hemispherical shell. A complete compensation of the systematic measurement uncertainty $\Delta m$ is thus possible by counter heating in the temperature range~$0.52<\tilde{\theta}_S<0.6$ for the considered Rayleigh number range, but the implementation in real measurements is difficult. Therefore, geometry variations with a constant temperature specification of $\tilde{\theta}_P=\tilde{\theta}_S=1$ are performed in the following. 

Among the geometric quantities investigated are the plate spacing $\tilde{a}$ and the spherical shell radius $\tilde{r}_S$. Figure \ref{img:BeeinAbstand}a demonstrates the reduction of the mass difference due to the additional baffle plate in relation to the reference values of free convection without additional built-ins. The greatest effect is achieved at a distance of $\tilde{a} = 1$, while the differences between $\tilde{a}=2$ and $\tilde{a}=3$ are small in the entire Rayleigh number range. The reasons for this progression are the stronger influence of the pressure increase on the underside of the plate and the altered flow pattern, which results in a detachment of the boundary layers. 

The negative values of the apparent mass difference for the flow suppression by a hemispherical shell in Fig. \ref{img:BeeinAbstand}b indicate a reversal of direction in the resulting updraft force. This trend can be explained by the strong slowdown of the flow and the pressure increase above the mass prototype. In accordance with the result of the straight baffle plate, these phenomena have the greatest impact for the smallest investigated radius $\tilde{r}_S$. As the Rayleigh number rises, buoyancy forces are more pronounced, causing the pressure gradient under the hemispherical shell and thus the resulting pressure force to increase in magnitude. Consequently, decreasing $\Delta m$ are visible in Fig.~\ref{img:BeeinAbstand}b.

In conclusion, the most effective of all discussed ways to alter the updraft force is provided by the hemispherical shell as a built-in with counter heating. Nevertheless, a full compensation of the systematic measurement deviation cannot be achieved with the variation of the radius $\tilde{r}_S$ at $\tilde{\theta}_S=1$ since there is no change in sign in the observed interval. Further considerations to limit the hemispherical shell in circumferential direction are planned at this point. An adaptive lifting and lowering of the obstacle depending on the existing temperature difference is conceivable in the real operation of the mass comparator.

\section{Three-dimensional analysis}
\label{3Dresults}
\subsection{Free convection without built-ins}
\begin{figure}[h]
	\centering
	\includegraphics[scale=0.6]{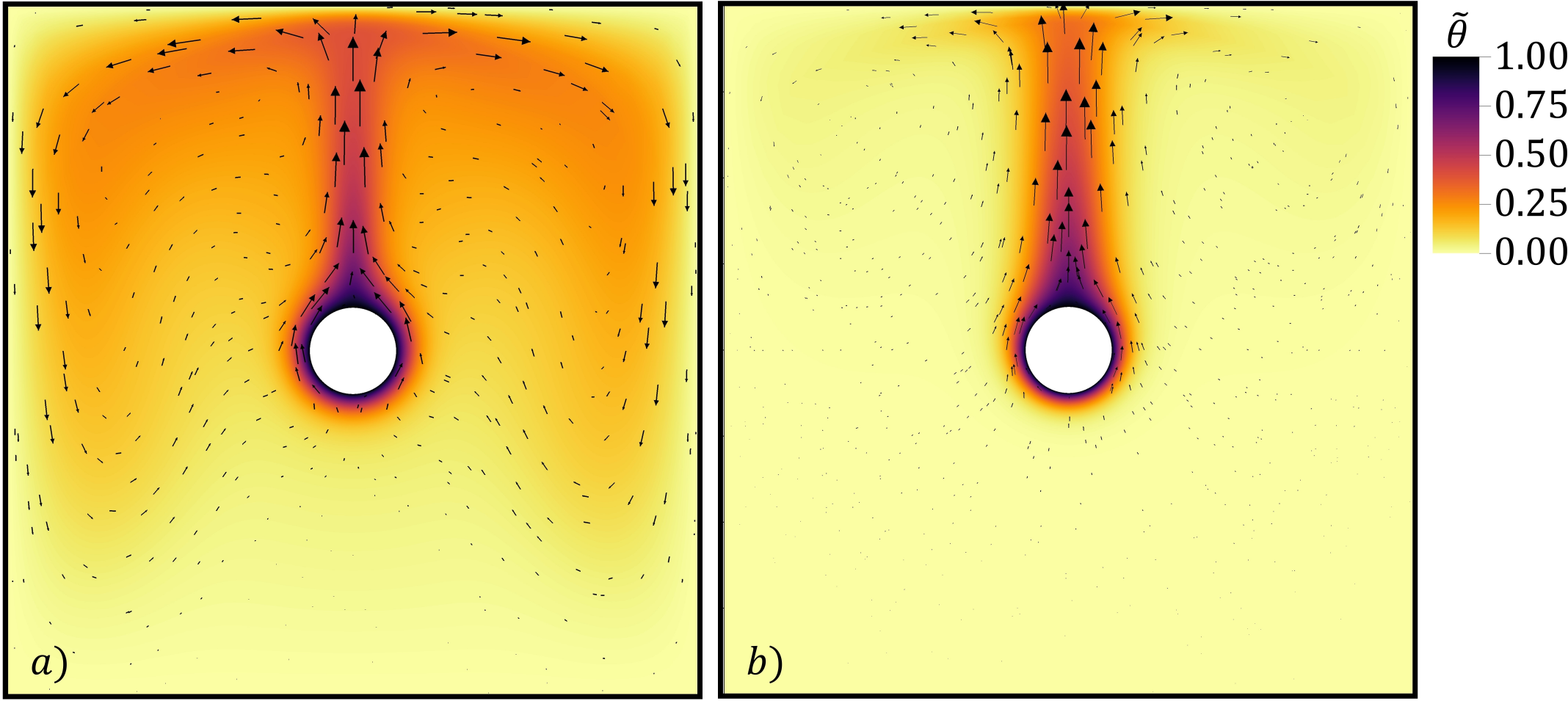}
	\caption{Comparison of the steady-state temperature fields as contour plots and the velocity distribution as vector plots for the 2d (a) and 3d (b) calculation at a Rayleigh number of $Ra=4590$ and a free-fall time of $\tilde{t} = 150$.}
	\label{img:temp3D}
\end{figure}

The two-dimensional analysis is extended in the following to the fully three-dimensional flow problem. First the free convection around the spherical mass standard, which is positioned  in the center of a cubical computational domain with a side length $\tilde{k}=8$, is discussed. At the outer surfaces of the fluid space no-slip boundary conditions in combination with a constant temperature $\tilde{\theta}=0$ are applied. A comparison of the steady-state temperature and velocity distribution is shown in Fig. \ref{img:temp3D} for a Rayleigh number of $Ra=4590$. The three-dimensional simulation results are discussed mostly in one of the symmetry planes, the $x$-$z$-plane.

The heated fluid rises now over the entire surface of the mass standard and initiates a flow with a three-dimensional character, which is not considered in the 2d models. The limitations of the two-dimensional case become particularly visible in the region of recirculation, which is significantly more pronounced compared to the 3d case. These have a decisive influence on the temperature distribution within the chamber and thus also affect the formation of boundary layer flows.
\begin{figure}[h]
	\centering
	\includegraphics[scale=0.4]{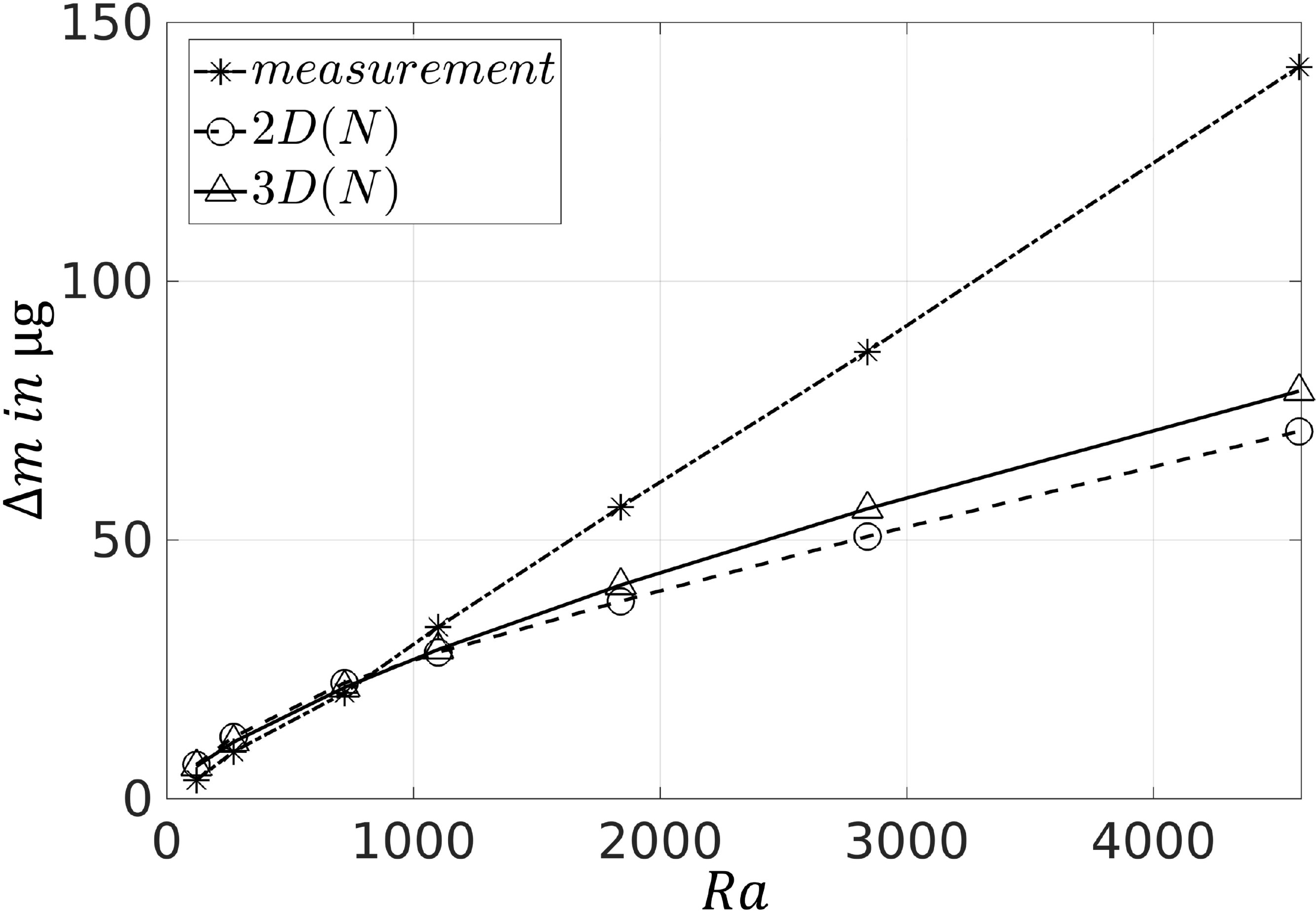}
	\caption{Numerical results (N) of the mass difference $\Delta m$ obtained in 2d and 3d calculations. These results are again compared to the measurements of Schreiber et al. \cite{Schreiber2015}. The side length of the domain is $\tilde{k}=8$.}
	\label{img:unbeein3D}
\end{figure}

A comparison of the resulting mass difference $\Delta m$ is displayed in Fig.~\ref{img:unbeein3D}. The data are also compared  to the measurements by Schreiber et al. \cite{Schreiber2015} for the Rayleigh number range. Starting from almost identical values for small Rayleigh numbers, the deviations between the numerically determined curves (N) increase. In the 3d calculation, larger values of the updraft force are obtained for all $Ra$. They are mainly a result of a stronger pressure increase due to the stagnation point formation at the south pole and the weaker pressure recovery at the north pole. Both effects lead to an increase of the pressure component of the updraft force. The consequences of recirculation are thus particularly evident at large Rayleigh numbers. Overall, the differences between both simulation series are rather small. Nevertheless, an additional approximation to the measured values could be achieved on the basis of the three-dimensional simulations.

\subsection{Secondary flow structure}
\begin{figure}[h]
	\centering
	\includegraphics[scale=0.51]{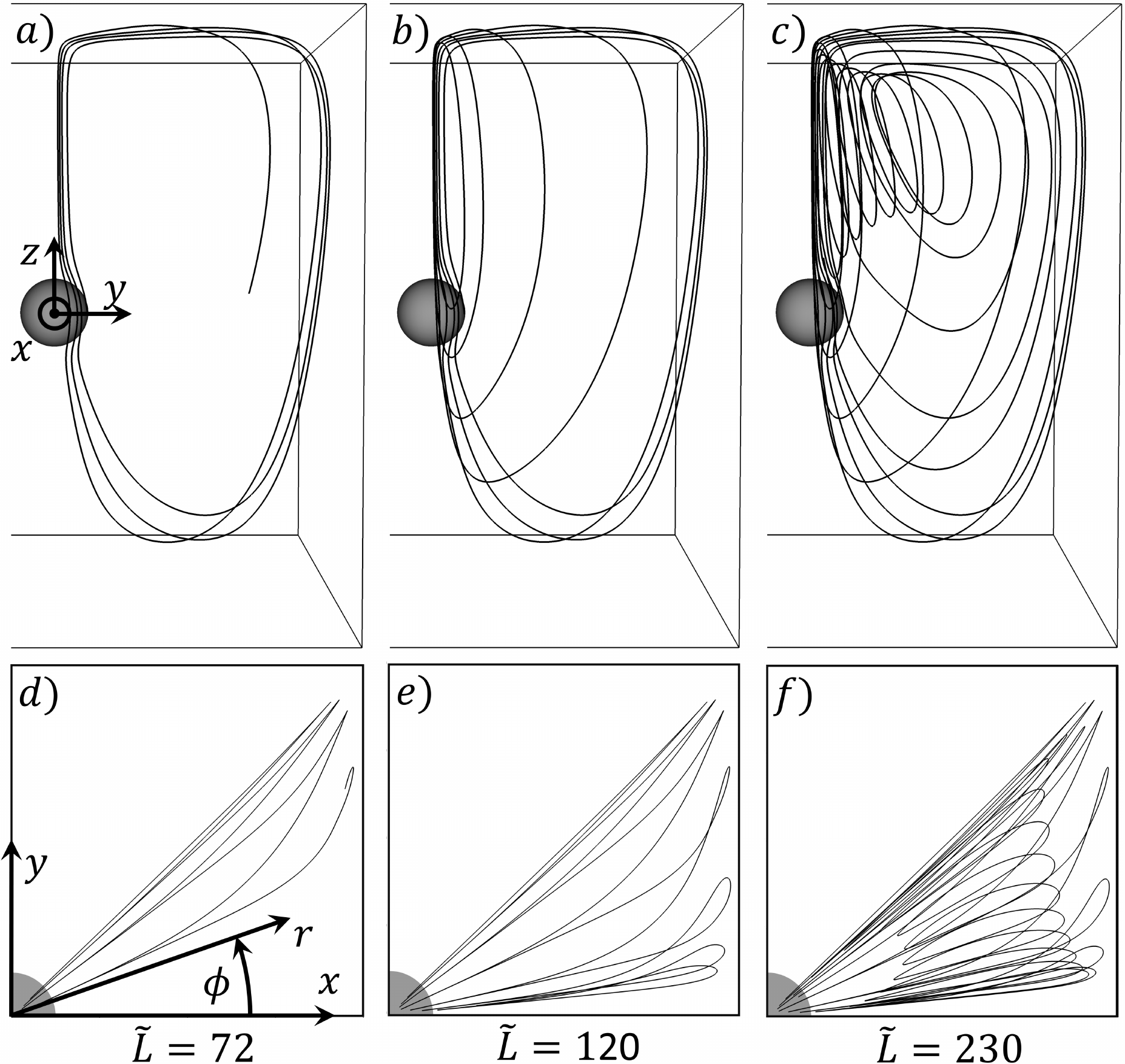}
	\caption{Three-dimensional side view (a,b,c) and top view (d,e,f) of a single streamline with increasing length~$\tilde{L}$, starting at an azimuthal angle of $\phi=44^\circ$. Data are for a Rayleigh number $Ra=4590$ at a time $\tilde{t}=150$ in a cube with side length  $\tilde{k}=8$.}
	\label{img:sekundaer}
\end{figure}

The 3d simulations reveal a complex structure of the secondary flows that occurs in the cubic chamber even at these small Rayleigh numbers. We therefore visualize an individual streamline stroboscopically with successively longer lengths $\tilde{L}$ to visualize the three-dimensional character of the secondary circulations of the velocity field better. This streamline is plotted in panels (a,b,c) of Fig.~\ref{img:sekundaer} in a side view perspective. The same streamline is again shown in panels (d,e,f), now from a top view perspective. Note that 8 such streamlines exist, 2 mirror-symmetric pairs in each of the four quadrants of the cube. Recall that each streamline has to be closed eventually due to flow incompressibility.

Beside the circulation flow there is a horizontal drift of the streamline towards the $x$-$z$-plane, which is observable when comparing Figs. \ref{img:sekundaer}a and \ref{img:sekundaer}b or Figs. \ref{img:sekundaer}d and \ref{img:sekundaer}e. This development has its origin in the flow deflection at the lateral boundaries. According to Figs. \ref{img:sekundaer}d and \ref{img:sekundaer}e, the fluid in the upper region of the domain gets initially diverted in the direction of the corner (i.e. along $\phi=45^\circ$), causing the pressure to increase locally. Taking into account the conservation of mass, the fluid then flows towards the $x$-$z$-plane in form of recirculations at lower velocity magnitudes. At the same time the streamline approaches the primary vortex center. In the core of this vortex, the fluid circulates only in the upper region of the domain, causing a deflection of the streamline primarily towards the corner of the fluid space. Thus the secondary flow is closed in Figs. \ref{img:sekundaer}c and \ref{img:sekundaer}f. Such streamline structure cannot be observed when starting the streamline in the four planes with an angle $\phi=n \times 45^\circ$ with $n\in\mathbb{N}$, because the main flow is only deflected in vertical direction at the lateral boundaries. We note finally that these kind of secondary flow features would also not be observable in an axisymmetric setup.

\subsection{Onset of unsteady convection}
An increase of the Rayleigh number to $Ra\geq2.5\times 10^4$ leads to the onset of a significant time dependence of the velocity and temperature fields in the two-dimensional simulations. These are initially characterized by lateral oscillations of the flow column and transform for $Ra \geq 10^5$ into periodic vortex shedding directly above the mass standard. To resolve the filamented vortex structures, the number of spectral elements was increased to 12800. With the three-dimensional modeling of the flow problem, stationary flow states were still observed at $Ra\sim 10^5$. Only for $Ra \gtrsim 10^6$, individual vortex structures in the region of the flow deflection at the upper boundary appeared, which increasingly influence the flow column. For these calculations the number of spectral elements was set to 92160 and the polynomial order of the basic functions to $N=11$. Since the focus of the present work is on the real and much smaller temperature differences of the real experiment, these investigations were not further detailed and are discussed here in brief only for completeness.

\subsection{Flow suppression by a hemispherical shell}
\begin{figure}[h]
	\centering
	\includegraphics[scale=0.38]{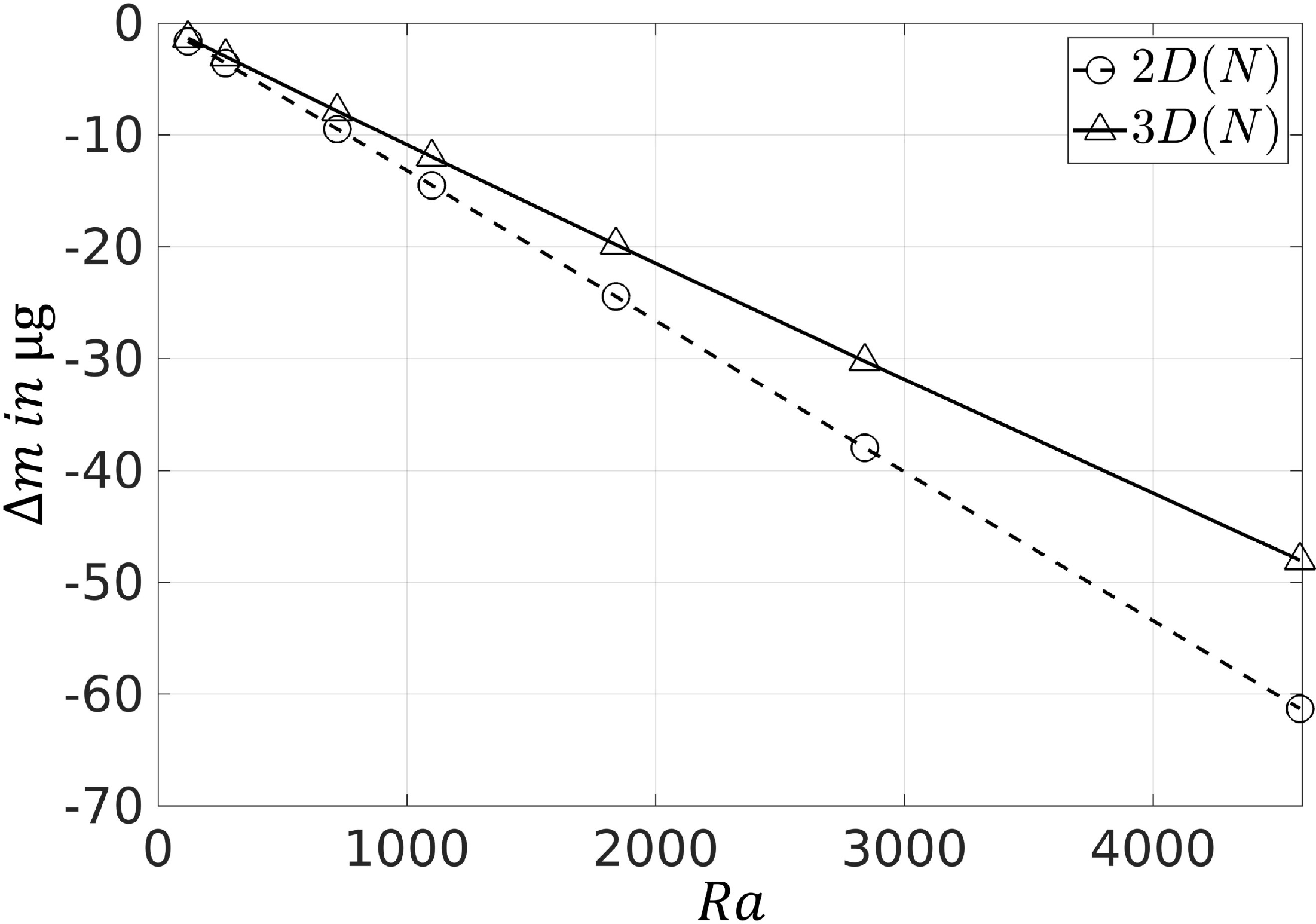}
	\caption{Mass difference in the three-dimensional simulations in comparison to corresponding 2d runs. In all cases a hemispherical shell with $\tilde{r}_S=1$ and $\tilde{\theta}_S=1$ was positioned above the spherical mass standard.}
	\label{img:beein3D}
\end{figure}

Using the modular mesh structure according to section \ref{mesh}, the targeted flow suppression by an additional hemispherical shell with a radius of $\tilde{r}_S=1$, no-slip boundary conditions and a surface temperature specification of $\tilde{\theta}_S=1$ is investigated finally in the three-dimensional case. A discussion of the mass difference is possible on the basis of Fig. \ref{img:beein3D}. As a result of the limited flow around the body and the pressure gradient below the obstacle, the negative values confirm a reversal of the updraft force for the 3d case. Differences to the planar model are found to grow with the Rayleigh number, since the flow organisation in three dimensions can significantly differ from that of the planar 2d case.  The physical mechanisms described in section \ref{Plate} have a smaller influence, which means that larger magnitudes of $\Delta m$ are always calculated in the 3d case. As a consequence of the weaker recirculation, convection processes in the immediate vicinity of the mass prototype are more developed and influence the pressure distribution. Consequently, the frictional component of the updraft force increases, while the magnitude of the negative pressure component decreases. In summary, the three-dimensional calculations confirm the possible compensation effect of the additional hemispherical shell above the mass standard. The evaluation of the apparent mass difference demonstrates a reversal of the updraft force which is, however, less pronounced as in the planar case. 

\begin{figure}[h]
	\centering
	\includegraphics[scale=0.38]{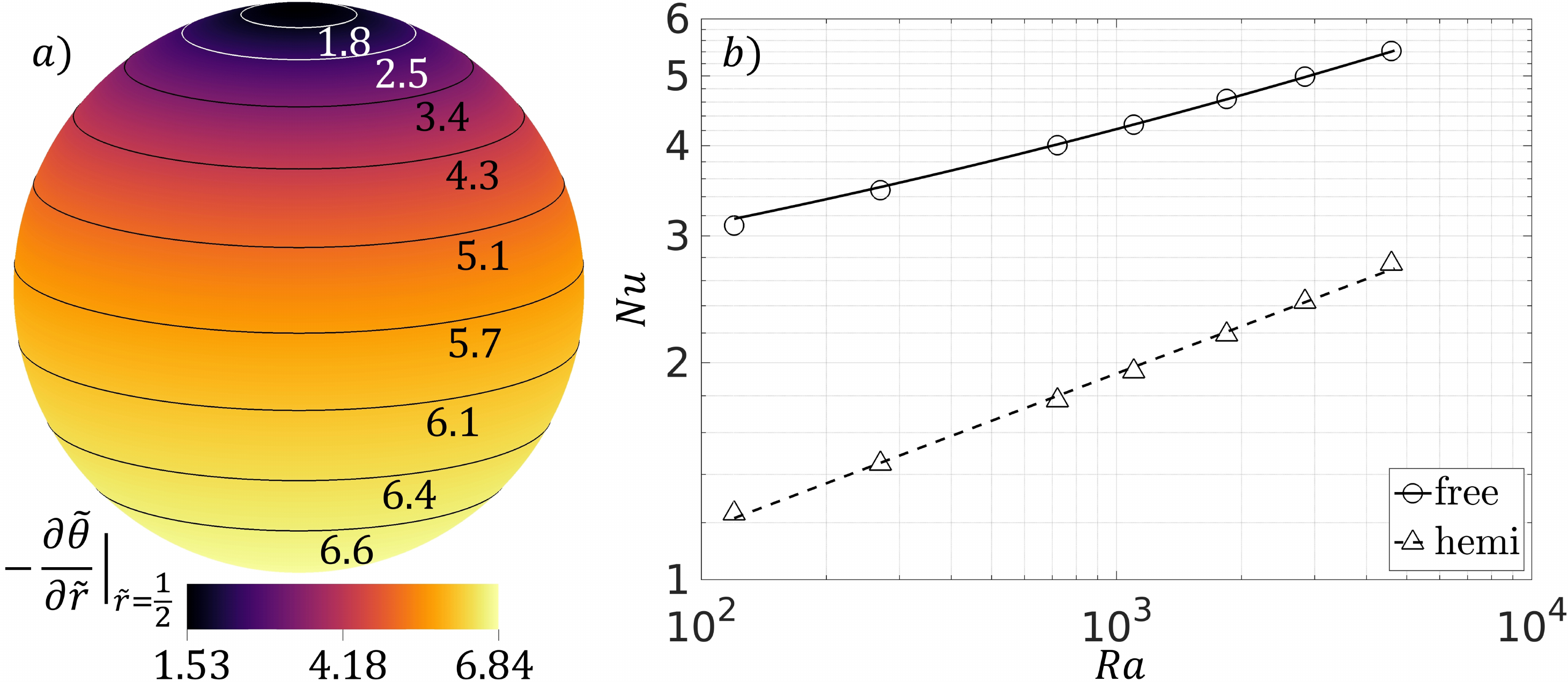}
	\caption{Heat transfer analysis in the three-dimensional simulations. (a) Distribution of the local Nusselt number across the surface of the spherical mass standard. See the legend. (b) Power laws of $Nu(Ra)$ for the case without built-ins and for the case with a hemispherical shell. The solid line stands for $Nu_{\rm free}=1.66+0.45\times Ra^{0.25}$ and the dashed one $Nu_{\rm hemi}=0.43 \times Ra^{0.22}$.} 
	\label{img:Nusselt}
\end{figure}

The free convection dynamics in our flow setup is in line with a heat transfer from the mass standard to the outer boundaries and the built-ins (in case of $\tilde{\theta}_P>0$ or $\tilde{\theta}_S>0$). The dimensionless measure of the heat transfer is given by the Nusselt number which is defined as
\begin{equation}
Nu=-\Bigg\langle\frac{\partial \tilde{\theta}}{\partial \tilde{r}}\Bigg|_{\tilde r= \frac{1}{2}}\Bigg\rangle_{A_s}\,.
\end{equation}
We thus average the diffusive heat at the surface $A_s$ of the spherical mass standard which is calculated in the spectral element model similarly to the resulting updraft forces. Panel (a) of Fig. \ref{img:Nusselt} shows the spatial distribution of the local Nusselt number over the sphere. Panel (b) of the same figure displays the relations $Nu(Ra)$ which result for the three-dimensional cases without built-ins (free) and with the hemispherical shell (hemi). Following \cite{Churchill1983,Incropera}, the function $Nu(Ra,Pr)$ for a sphere in free space is given by the following empirical law (for $Pr\ge 0.7$ and $Ra\le 10^{11}$)
\begin{equation}
Nu_{\rm free}=2+ \dfrac{0.589\, Ra^{1/4}}{\left[1+\left(\dfrac{0.469}{Pr}\right)^{9/16}\right]^{4/9} }\,.
\label{rel1}
\end{equation}
Our data for the free and hemispherical cases could be fitted by the following functions (here $Pr=0.71$), 
\begin{equation}
Nu_{\rm free}=1.66+0.45\times Ra^{1/4} \quad\mbox{and} \quad Nu_{\rm hemi}=0.43 \times Ra^{0.22}\,.
\end{equation}
The constant offset in the free case is reduced from 2 in \eqref{rel1} to 1.66 which we attribute to the backflow in the closed chamber that reduces the heat transfer slightly.  The other coefficients of \eqref{rel1} were left unchanged. It can also be seen that the heat transfer is significantly suppressed for the case with the hemispherical shell and remains close to lower diffusive bound of $Nu=1$.

\section{Conclusion and outlook}
\label{Conclusion}
The mass difference resulting from free convection during the weighing of spherical 1 kg mass standards (or prototypes) was investigated in relation to previous measurements by Schreiber et al. \cite{Schreiber2015} by direct numerical simulations with the spectral element solver Nek5000. A good quantitative agreement was obtained for the smaller temperature differences up to 15mK which corresponds to $Ra \lesssim 1200$, whereas the deviations increase with growing Rayleigh number. First, we analysed the laminar boundary layer structure around the sphere and investigated the impact of different outer domain boundary conditions of the velocity field and the domain size on the mass difference in a series of 2d simulations. Outflow boundary conditions are found to enhance the mass difference by about 14\%. Secondly, we studied the effect of additional built-ins and of their position and extension in the chamber on the mass difference. Two geometries, which are positioned above the mass standard, were suggested here -- a straight baffle plate and a hemispherical shell. Particularly, the application of the latter one made a full compensation of $\Delta m$ (and thus the updraft forces due to free convection) possible. This step demonstrated that the resulting effective forces can be suppressed by appropriate counter heating of these built-ins.

On the basis of subsequent three-dimensional calculations the differences to the measurements could be further reduced slightly. Remaining deviations exist to our view mainly due to the geometric simplification of the measurement chamber. A further inclusion of other devices (such as the pans that hold the mass standard) into the numerical study would provide a direction of possible future work. More complex simulations that capture the influence of the additional components of the mass comparator are thus useful and can provide the key to an improved prediction of the apparent mass difference. Furthermore, the specification of a constant surface temperature at the mass standard is a simplifying assumption that would have to be tackled by additional solid state grids for the sphere. This would imply to convert the problem into one with conjugate heat transfer. Finally, it can be expected that the temperature differences will remain in the lower-Rayleigh-number range in a precision measurement setup where the agreement between measurement and simulation was very well.

As already mentioned, the modular extension of the mesh allowed the targeted flow suppression by a straight baffle plate and a hemispherical shell, each positioned above the mass standard. One result of  our study is a reduction of the convective processes by at least $65~\%$ which can be achieved with the baffle plate at a spacing $\tilde{a}=1$ and for the constant surface temperature $\tilde{\theta}_S=1$. When using the hemispherical shell, a reversal of the updraft force was shown, which could be confirmed in three-dimensional simulations. A complete compensation of the systematic measurement uncertainty is thus conceivable by varying the extension of the obstacle in circumferential direction. The corresponding adaptation of the grid is planned in the next step. For the practical implementation in high-precision measuring devices, mechanisms for lifting and lowering the additional obstacle depending on the existing temperature difference would be possible.

\appendix

\section{Sensitivity of the two-dimensional Mesh}
\label{Sensitivity}
The spatial resolution of the grid increases significantly with the number of elements and the polynomial order of the basis functions and is of central importance for the quality of the numerical solution. Table \ref{tab:Elementanzahl} summarizes a grid sensitivity study. Here, an increase in the number of time steps as well as in the computation time becomes visible, if the number of spectral elements in Fig. \ref{img:Mesh}a is increased in five successive levels. The polynomial order of the basis functions remains constant at~$N=7$ for all cases.
\begin{table}[h]
	\centering
	\begin{tabular}{llllll}
		\toprule[1pt]
		level & 1 & 2 & 3 & 4 & 5\\
		\midrule[1pt]
		number of elements &  & & & & \\
		radial direction & 10 & 20 & 40 & 80 & 120 \\
		circumference direction & 20 & 40 & 80 & 160 & 240 \\
		total number & 200 & 800 & 3,200 & 12,800 & 28,800 \\
		number of time steps & 4,102 & 8,349 & 17,901 & 36,698 & 51,366 \\
		computation time in s & 18 & 55 & 470 & 12,372 & 58,128 \\
				
		\bottomrule[1pt]
		
	\end{tabular}
	\caption{Number of elements and iterations with the required computation time for simulations of $\tilde{t}=30$ on 48 cores for $Ra=1,500$, $\tilde{k}=12$ and $N=7$. }
	\label{tab:Elementanzahl}
\end{table} 

\begin{figure}[h]
	\centering
	\includegraphics[scale=0.46]{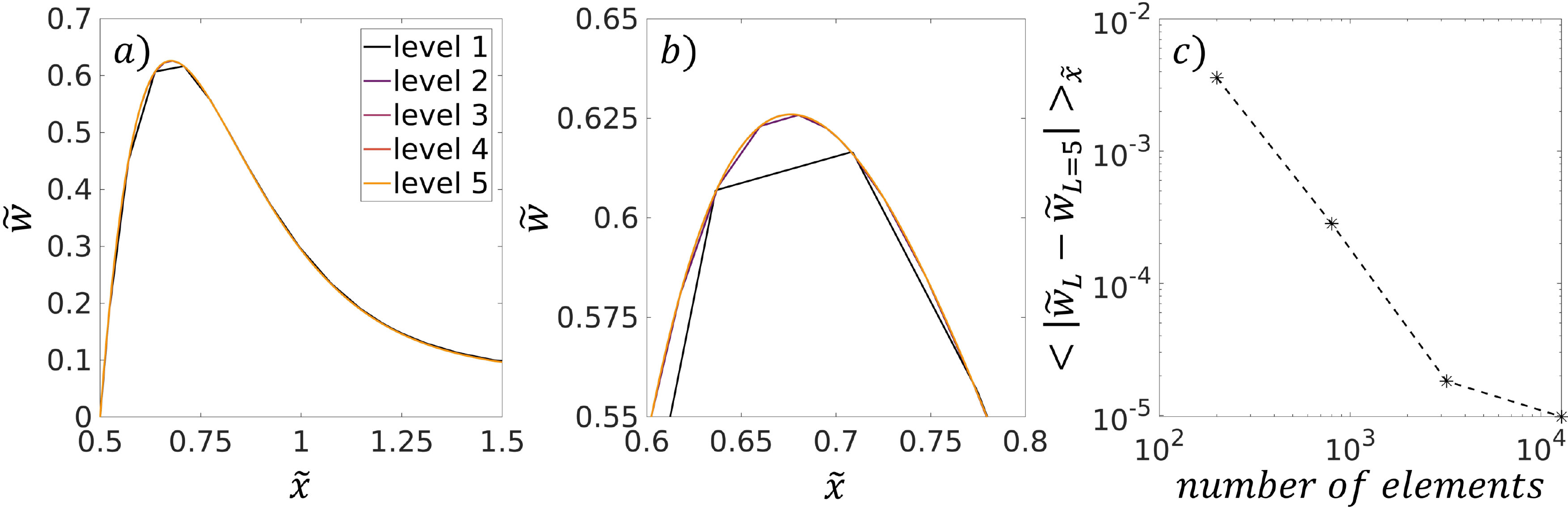}
	\caption{Velocity boundary layer profile $\tilde{w}(\tilde{x})$ at $\tilde{z}=0$ in the interval $\tilde{x}\in[0.5,1.5]$ (a) and $\tilde{x}\in[0.6,0.8]$ (b). Mean value of the deviations from refinement with level 5 (c). We have $Ra=1,500$, $\tilde{k}=12$, $\tilde{t}=30$, and $N=7$.}
	\label{img:Meshelem}
\end{figure}
To assess the quality of the results, velocity profiles $\tilde{w}(\tilde{x})$ within the boundary layer at $\tilde{z}=0$ are compared in Figs. \ref{img:Meshelem}a and \ref{img:Meshelem}b. Deviations occur near the maximum especially for the levels 1 and 2, while the curves are almost identical for the levels 3 to 5. In Fig. \ref{img:Meshelem}c the arithmetic mean of the deviations from the velocity profile with the highest mesh refinement is plotted over the number of elements. While a rapid decrease is reflected between levels 1 to 3, the curve flattens out considerably at level~4. An independence of the number of elements is thus achieved in good approximation from level 3 on. Taking the computational effort into account, the refinement with 3,200 elements was used in the simulations.
\begin{table}[h]
	\centering
	\begin{tabular}{llllll}
		\toprule[1pt]
		polynomial order & 3 & 5 & 7 & 9 & 11 \\
		number of time steps & 4,337 & 10,193 & 17,901 & 29,515 & 41,619 \\
		computation time in s & 36 & 157 & 470 & 1,307 & 3,103 \\
		\bottomrule[1pt]
		
	\end{tabular}
	\caption{Number of iterations with the required computing time for simulations with $Ra=1,500$, $\tilde{k}=12$ and $3,200$ elements depending on the polynomial order. }
	\label{tab:Polynomordnung}
\end{table} 

\begin{figure}[h]
	\centering
	\includegraphics[scale=0.46]{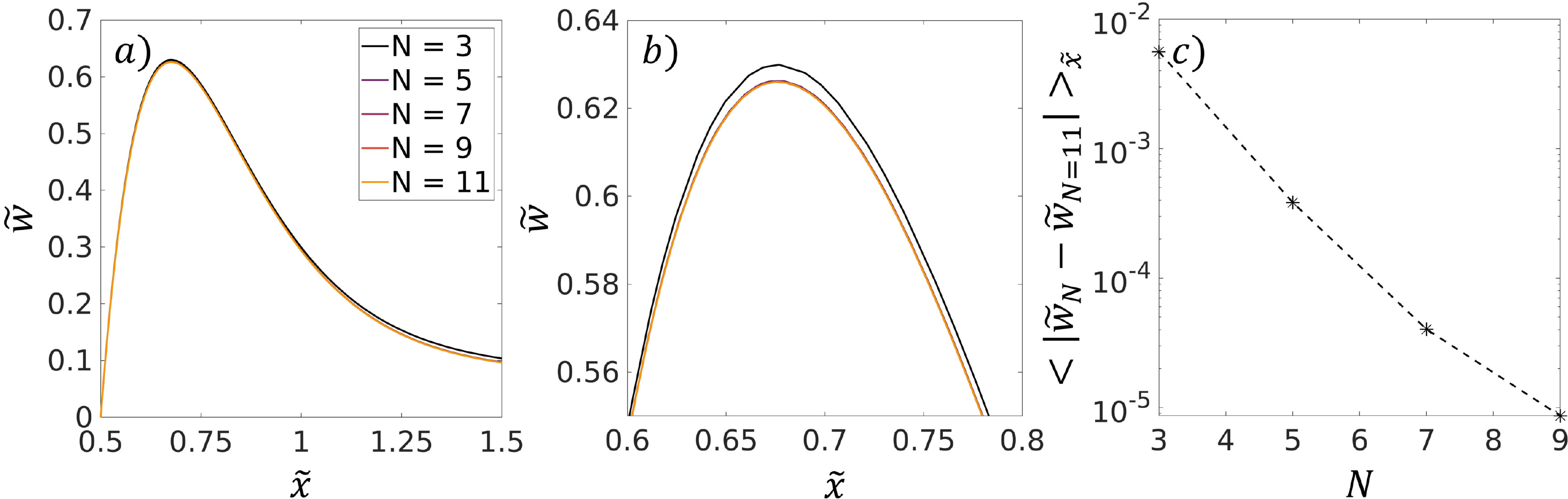}
	\caption{Velocity boundary layer profile $\tilde{w}(\tilde{x})$ at $\tilde{z}=0$ in the interval $\tilde{x}\in[0.5,1.5]$ (a) and $\tilde{x}\in[0.6,0.8]$ (b). Mean value of the deviations from refinement with the polynomial order $N=11$ (c). We have $Ra=1,500$, $\tilde{k}=12$, and $\tilde{t}=30$. In the cases, $3,200$ spectral elements are used.}
	\label{img:Meshpoly}
\end{figure}
With the specified number of elements identical calculations were performed with a varying polynomial order $N$ on each spectral element and in each space direction according to Table \ref{tab:Polynomordnung}. With the refinement of the mesh, an increase of the computational effort is again observed. Increasingly smoother curves are shown in a comparison of the velocity profiles $\tilde{w}(\tilde{x})$ in Fig. \ref{img:Meshpoly}a and \ref{img:Meshpoly}b. Significant deviations are only visible for $N=3$. Figure \ref{img:Meshpoly}c shows an exponential convergence with increasing polynomial order, which is typical for the spectral element method. With regard to the quality of the results and the computing time, the polynomial order of the basic functions in the simulations was set to $N=7$.

\section{Modular mesh expansion}
\label{Modular}
\begin{figure}[h]
	\centering
	\includegraphics[scale=0.46]{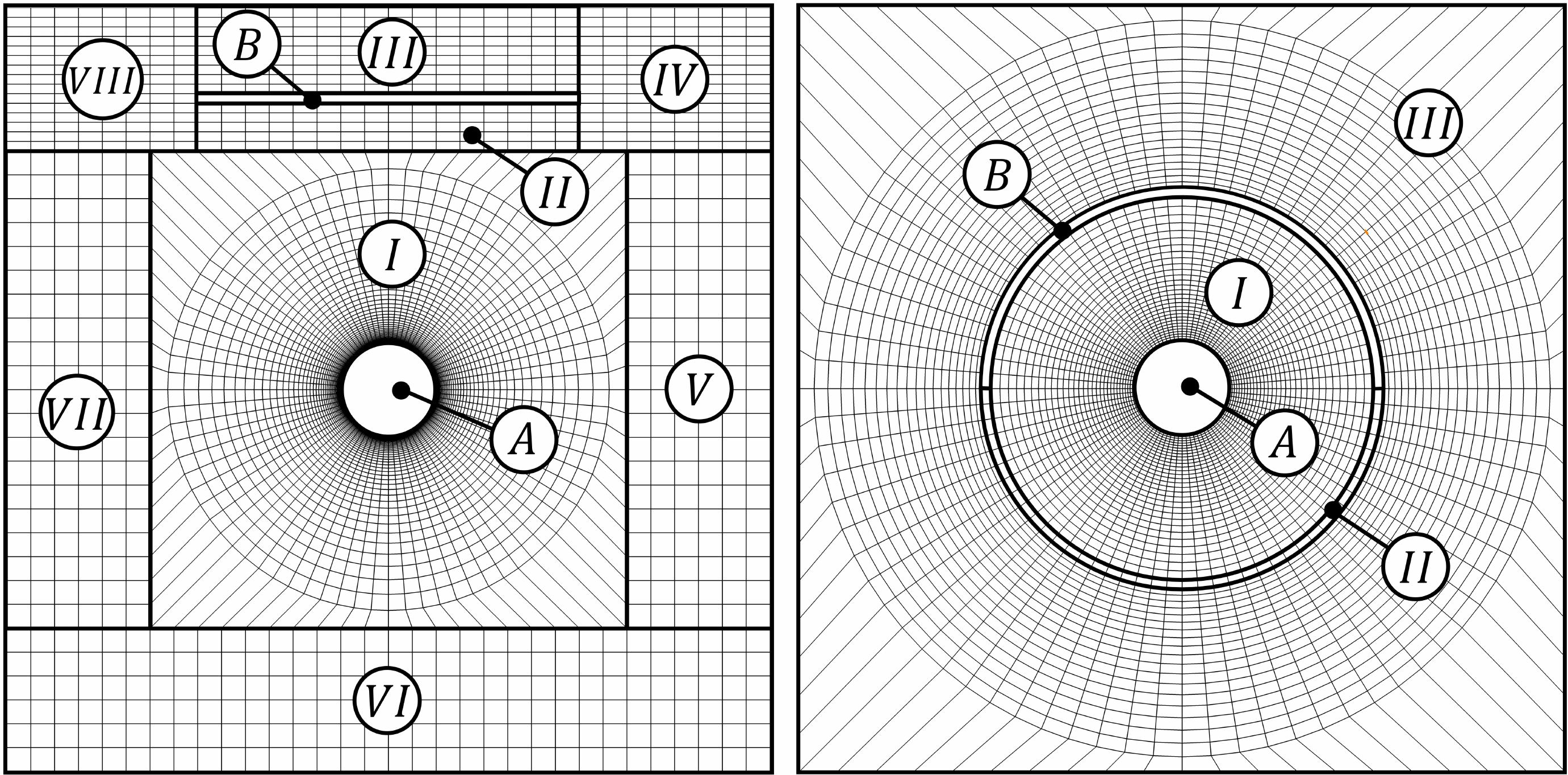}
	\caption{Modularly extended mesh with the subgrids $I-VIII$ for the straight baffle plate (left) and $I-III$ for the hemispherical shell (right). The mass standard (A) and the obstacles (B) were modeled as free space.}
	\label{img:Meshprall}
\end{figure}

In order to model the additional obstacles in the flow path, the cylinder mesh is extended in a modular way. The individual subgrids are first discretised separately and then merged together. Similar to the mass standard, the obstacles are modeled by a free space with no-slip boundary conditions. Figure \ref{img:Meshprall} shows the mesh structure for the configurations introduced in section \ref{mesh}. In the case of the planar baffle plate eight subgrids ($I-VIII$) are combined, whereas only three ($I-III$) are necessary for the hemispherical shell. For the mesh refinement, additional elements were implemented instead of the obstacles and the results for the unrestricted flow were compared with those from section \ref{BC}. Finally, the number of elements was increased until the maximum uncertainty regarding the apparent mass difference was in the range of $\pm 0.02~\rm{\mu g}$.

The modeling of the hemispherical shell in the three-dimensional case is analogous to the presented extension of the cylinder mesh. Internal boundary conditions are automatically set at the connections of the partial meshes to ensure that the calculated flow fields match for adjacent elements.

\end{document}